\documentclass [11pt]{article}
 \pdfoutput=1
 
  \usepackage{axodraw4j}
\usepackage{pstricks}

 \usepackage{tikz}
\usetikzlibrary{decorations.pathmorphing}
\usetikzlibrary{patterns}
\usetikzlibrary{cd}
\usetikzlibrary{snakes}
\usepackage{latexsym,amsmath,amsfonts,amssymb,amsthm}
\usepackage{mathrsfs}
\usepackage{enumitem}
\usepackage[latin1]{inputenc}
\usepackage[american]{babel}
\usepackage{bbm}
\usepackage{cite}
\usepackage{amsfonts}
\usepackage{graphicx}
\usepackage{amsmath,bm}
\usepackage{amssymb}
\usepackage{latexsym}
\usepackage{mathrsfs}
\usepackage{slashed}
\definecolor{red}{rgb}{1,0,0}
\usepackage{cancel}

\usepackage{wasysym}
\usepackage{soul} 

\usepackage{setspace} 

\theoremstyle{plain}
\newtheorem{thm}{Theorem}[section]

\theoremstyle{definition}
\newtheorem{defn}[thm]{Definition}

\theoremstyle{remark}

\newtheorem{exr}{$\mathparagraph$ Exercise}

\newlist{numlist}{enumerate}{1}
\setlist[numlist]{label=(\roman*), topsep=0.2em, itemsep=0em}

\newlist{alphlist}{enumerate}{1}
\setlist[alphlist]{label=(\alph*), topsep=0.2em, itemsep=0em}

\definecolor{red}{rgb}{1,0,0}

\makeatletter
\def\section{\@startsection {section}{1}{\z@}{-3.5ex plus -1ex minus
 -.2ex}{2.3ex plus .2ex}{\large\bf}}
\def\subsection{\@startsection{subsection}{2}{\z@}{-3.25ex plus -1ex
minus -.2ex}{1.5ex plus .2ex}{\normalsize\bf}}
\makeatother
\makeatletter

\@addtoreset{equation}{section}

\makeatother

\def\bel{\begin{equation}\begin{aligned}}
\def\eel{\end{aligned}\end{equation}}

\def\bea{\begin{eqnarray}} \def\eea{\end{eqnarray}}
\def\be{\begin{equation}} \def\ee{\end{equation}} 
\def\nn{\nonumber}

\newcommand{\promille}{%
  \relax\ifmmode\promillezeichen
        \else\leavevmode\(\mathsurround=0pt\promillezeichen\)\fi}
\newcommand{\promillezeichen}{%
  \kern-.05em%
  \raise.5ex\hbox{\the\scriptfont0 0}%
  \kern-.15em/\kern-.15em%
  \lower.25ex\hbox{\the\scriptfont0 00}}

\usepackage[colorlinks,linkcolor=black,citecolor=blue,urlcolor=blue,linktocpage]{hyperref}

\begin{document}
\setcounter{page}{0}
\thispagestyle{empty}

\parskip 3pt

\font\mini=cmr10 at 2pt

\begin{titlepage}
~\vspace{1cm}
\begin{center}

{\LARGE \bf Lectures on Resurgence}\\[4mm]
{\LARGE\bf  in Integrable Field Theories}

\vspace{1.6cm}

{\large
Marco~Serone}
\\
\vspace{1cm}
{\normalsize { \sl  
SISSA International School for Advanced Studies and INFN Trieste, \\
Via Bonomea 265, 34136, Trieste, Italy }}

\end{center}

\begin{abstract}

There has been recently considerable progress in understanding the nature of perturbation theory in UV free and gapped $2d$ integrable field theories with renormalon singularities.
Thanks to Bethe ansatz and large $N$ techniques, non-perturbative corrections can also be computed and lead to the reconstruction of the trans-series for the free energy in presence of a chemical potential. This is an ideal arena to test resurgence in QFT and determine if and how the exact result can be reconstructed from the knowledge of the perturbative series only. In these notes we give a pedagogical introduction to this subject starting from the basics. In the first lecture we give an overview of applications in QFT of Borel resummations before
the advent of resurgence. The second lecture introduces the key concepts of resurgence and finally in the third lecture we discuss a specific application in the context of the principal chiral field model. Extended version of three lectures given at IHES and review talks given at Les Diablerets and Mainz, in 2023.

\end{abstract}

\end{titlepage}

\tableofcontents

\section{Introduction}

Perturbation theory is one of the most successful analytical tools to study physical processes and is hence very important to understand its nature.
It is known that perturbative expansions in quantum field theories (QFT) are generically asymptotic with zero radius of convergence \cite{Dyson-1952}. 
In special cases, such as  $\phi^4$ theories up to $d=3$ space-time dimensions, the perturbative expansion
turns out to be Borel resummable \cite{Graffi:1970erh,Eckmann,Magnen}. 
Before the advent of resurgence in QFT, Borel summability was considered essential.
Early works  were mostly interested in the large order behaviour of the perturbative series, determined by certain saddle points in a complexification of the theory \cite{Lipatov:1976ny}.
This is useful to numerically reconstruct the Borel function from its first few known perturbative coefficients, giving in this way an estimate of the observable.
A notable successful application of resummation methods along this line occurred in statistical physics in the determination of critical exponents of second order phase transitions.

Most interesting QFTs, such as gauge theories in four space-time dimensions, are however not Borel resummable because of singularities in 
the domain of integration of the Borel function. These can be avoided by deforming the contour 
at the cost of introducing an ambiguity that is non-perturbative in the expansion parameter. The singularities can be associated to real instantons, in which case 
we expect that the ambiguity might be removed by including them in the path integral, together with their corresponding series expansion, 
resulting in what is called  a trans-series.
A notable singularity which seems to be always present whenever the theory includes marginal couplings
is the so called renormalon singularity \cite{gross-neveu,Lautrup:1977hs,tHooft:1977xjm}. In contrast to instantons, we do not know the semi-classical configuration associated to renormalons (indeed it might simply
not exist).\footnote{As we will briefly discuss, renormalons can however be related to power corrections whenever the observable of interest admits an operator product expansion (OPE).}
Independently of its interpretation, the ambiguity resulting from the non-Borel summability of a series is due to Stokes phenomena. 
The theory of resurgence  \cite{ecalle} is a systematic way to deal with Stokes phenomena and trans-series, and give a meaning to otherwise non-Borel resummable asymptotic series.
According to resurgence, a quantity can be expanded in a trans-series which, as a whole,  will be ambiguity-free and Borel resummable.
Importantly, the asymptotic series entering in the trans-series are related to each other. In the best case scenario, one can hope that the whole trans-series can be reconstructed starting 
by the perturbative series only.

A natural arena where resurgence can be applied is in the study of differential equations, where (formal) solutions are expressed by asymptotic power series.  
Understanding the Stokes phenomena allows us to determine the actual well-defined solutions of the equations. 
A beautiful application of resurgence to the Schr\"odinger equation is at the base of the exact WKB method \cite{voros-quartic,Zinn-Justin:1982hva,AKT1,reshyper,ddpham,dpham,AKT2}.
Exact WKB is the upgrade of the WKB approximation to an exact method and allows us to get the explicit form of the
trans-series for wave functions, energy eigenvalues, etc. 

A natural question to ask is then: does resurgence work in QFT, in particular in theories featuring renormalon singularities?
This is a hard question in general, because in QFT  we do not have ``Schr\"odinger"- like equations to determine perturbative series at arbitrarily high orders.
However, there exist models which have renormalons and at the same time can be studied in detail, both in perturbation theory and in a $1/N$ expansion, 
namely integrable 2d asymptotically free theories. In these models both the spectrum and the S-matrix are known exactly and, as noted long ago \cite{pw},
one can use Thermodynamic Bethe Ansatz (TBA) techniques to compute exactly the free energy in presence of a chemical potential. 
Using also other results to extract perturbative series for the free energy at very high orders \cite{volin, volin-thesis}, there have been recently 
several studies to understand the resurgence nature of the perturbative expansion in integrable 2d theories \cite{mr-ren,abbh1,abbh2,mmr,DiPietro:2021yxb,Marino:2021dzn,Bajnok:2021dri,Bajnok:2021zjm,Bajnok:2022rtu,Marino:2022ykm,Bajnok:2022xgx,Schepers:2023dqk,Bajnok:2024qro}.

Aim of these lectures is to give a pedagogical introduction to large orders and resurgence aimed at an application in integrable models.
Aside technicalities, the key principles underlying resurgence are simple and we will focus on those.  We start in section \ref{sec:ASBR} from the basics, recalling why
we get asymptotic series in QFT and the conditions for Borel summability. We then give a quick historical excursus on the subject prior to resurgence, 
discussing how large orders are related to (complex) instantons, applications in critical phenomena, renormalons. 
We introduce resurgence in section \ref{sec:resurgence} by working out in detail the Airy function as a solution of a differential equation, and how
this is related to its integral representation. We then discuss a few formal  resurgence tools, the minimal ones required for our subsequent application.
In section \ref{sec:2d1N} we finally work out in detail a specific result found in \cite{DiPietro:2021yxb} where resurgence works at its best,  
the determination of the trans-series for the free energy at leading order in $1/N$ in the
principal chiral field model.  We will concretely show how the perturbative series is able to determine the exact non-perturbative result, summarized in our final relation \eqref{eq:TS13}.
A few concluding remarks are given in section \ref{conclusions}.

A comprehensive review on resurgence, including several examples, is \cite{abs}. A more compact, and slightly more math-oriented, review is \cite{Dorigoni:2014hea}, while
a systematic math-oriented review is \cite{sauzin2014}. See in particular \cite{abs} and references therein for an overview of topics in theoretical physics where resurgence has been applied.

\section{Asymptotic Series and Borel Resummation}
\label{sec:ASBR}

Let us start by reviewing well-known facts about elementary complex analysis. A function $f$ of a complex variable $g$ is 
 said to be analytic at a point $g_0$ if, in a small disc around $g_0$, the function is given by the power series
\be
f(g) =  \sum_{n=0}^\infty  c_n (g-g_0)^n\,.
\label{eq:ASBR1}
\ee
The radius of convergence $R$ of the series (\ref{eq:ASBR1}) is given by 
\be
R^{-1} = \limsup_{n\rightarrow\infty} |c_n|^{\frac 1n} \,.
\label{eq:ASBR2}
\ee
If $g_0$ is a regular point of $f$, then $R$ is non-vanishing and is given by the distance of  $g_0$ from the closest singularity of $f(g)$. Viceversa,
if $R$ is non-vanishing, necessarily $g_0$ is a regular point of the function $f$.
The power series (\ref{eq:ASBR1}) is uniformly convergent for any $|g|<R$ and divergent for $|g|>R$. The convergence for $|g|=R$ depends on the 
particular cases, but necessarily there is at least a point where the series diverges, corresponding to the singular point of $f(g)$ closest to $g_0$.

In 1952, using analyticity and physical arguments, F. Dyson argued that the perturbative series expansion in QED  has to have zero radius of convergence  \cite{Dyson-1952}. 
The basic argument is simple and brilliant. The QED expansion is a power series in $\alpha\propto e^2$, where $e$ is the electric charge.
Suppose that this expansion had a non-vanishing radius of convergence. Then the point $\alpha=0$ should be a regular point when we look at observables $O(\alpha)$
as analytic functions in $\alpha$. This would imply that $O(\alpha)$ has a finite radius of convergence where the function is analytic, including regions where 
$\alpha<0$.  But when $\alpha<0$ electrons and positrons repel each other and the vacuum is unstable. 
We conclude that physical observables in our world cannot be analytic at $\alpha=0$. In turn, this implies that the series around $\alpha=0$ 
has zero radius of convergence.

A modern version of Dyson's argument can be obtained by considering the euclidean path integral formulation in QFT.
Consider a generic $n$-point function of local operators in a theory described by an action $S$:
\be
G^{(n)}(x_1,x_2,\ldots x_n; \hbar) = \int\!{\cal D}\phi \,  \phi(x_1) \phi(x_2)\ldots \phi(x_n) \, e^{-S(\phi)/\hbar}\,,
\label{eq:ASBR3}
\ee
where we denote collectively all the fields in the action by $\phi$. In euclidean space the action is positive definite and the path integral converges.\footnote{With an appropriate measure and upon renormalization. Strictly speaking a full non-perturbative definition of the path integral
requires a lattice discretization of space-time.} Consider now the loopwise expansion, that is the expansion of $G^{(n)}$
in powers of $\hbar$. The point $\hbar=0$ is necessarily non-analytic because for any value of $\hbar <0$ the Green functions $G^{(n)}$ blow up. 
We conclude that loopwise perturbative expansions in QFT have generically zero radius of convergence and are divergent.
A similar conclusion applies for coupling constant expansions which, upon rescaling of the fields, are equivalent to a loopwise expansion.

But how well-defined physical observables give rise to asymptotic expansions? In order to answer to this question it is enough to consider ordinary integrals,
which we can interpret as baby version of path integrals in quantum mechanics or quantum field theory. Consider for example the integral
\be
I(g) =\int_{-\infty}^\infty \! dx\, e^{-\frac{x^2}{2}-\frac{gx^4}{4}} \,, 
\label{eq:ASBR4}
\ee
well-defined for real and positive $g$. Expanding for small $g$ we get
\be
I(g) =\int_{-\infty}^\infty \! dx\, e^{-\frac{x^2}{2}} \sum_{n=0}^\infty \frac{g^n}{n!}\Big(-\frac{x^4}{4}\Big)^n  \stackrel{?}{=}  
 \sum_{n=0}^\infty g^n \int_{-\infty}^\infty \! dx\, e^{-\frac{x^2}{2}}\frac{1}{n!}\Big(-\frac{x^4}{4}\Big)^n = \sum_{n=0}^\infty c_n g^n =  \widetilde I(g)\,,
\label{eq:ASBR5}
\ee
where
\be
c_n = \frac{\sqrt{2}(-1)^n \Gamma\Big(2n+\frac 12\Big)}{n!}   \underset{n\to\infty}{\longrightarrow}  n! (-4)^n\frac{1}{\sqrt{\pi}n} \Big(1+{\cal O}\big(\frac 1n\big)\Big)\,.
\label{eq:ASBR6}
\ee
The factorial growth of the coefficients $c_n$ implies that the series $\widetilde I(g)$ has zero radius of convergence and is asymptotic.
The mistake we made in \eqref{eq:ASBR5} is exchanging sum and integration.  Given a series of functions $\sum_n f_n(x)$ such that
\be\label{eq:ASBR7}
\sum_{n=0}^\infty \int_X |f_n(x)| < \infty \,,
\ee
then the dominated convergence theorem implies that (see e.g. \cite{Rudin})
\be\label{eq:ASBR8}
\sum_{n=0}^\infty \int_X f_n(x) =\int_X  \sum_{n=0}^\infty f_n(x)\,. 
\ee
The series in \eqref{eq:ASBR5} is convergent (the exponential) but it does not satisfy \eqref{eq:ASBR7}. So, out of a well-defined function $I(g)$, perturbation theory
would give us a formal ill-defined asymptotic series $\widetilde I(g)$. The same unjustified step is performed in QFT when we use perturbation theory.
Since path integrals do not make the situation better, perturbative expansions in QFT are generally asymptotic.

In order to not confuse a function with its formal asymptotic series, we denote by a tilde the formal asymptotic series of a given quantity $f(g)$. We write
\be\label{eq:ASBR9}
f(g)\sim \widetilde{f}(g)
\ee
to indicate that $f$ has $\widetilde{f}$ as formal asymptotic series. 

A series $\widetilde f(g) = \sum_{n=0}^\infty c_n g^n$ is denoted asymptotic if, for any fixed order $N$, we have 
\be
f(g) - \sum_{n=0}^{N-1} c_n g^n = {\cal O}( g^{N}) \,, \quad {\rm as} \quad  g\rightarrow 0\,.
\label{eq:ASBR10}
\ee
Note the crucial difference with respect to convergent series where for $N\rightarrow \infty$ the sum $\sum_{n=0}^N c_n  g^n$ 
approaches $f(g)$ for {\it any} $ g$ within the domain of convergence.
For convergent series different functions lead to different series. This is not the case for asymptotic expansions, where
different functions can have the same asymptotic expansion. For example, two functions $f_1( g)$ and $f_2(g)$ with
\be
f_2(g) = f_1( g) + e^{-1/ g} a( g)\,,
\label{eq:ASBR11}
\ee
have the same asymptotic expansion, provided $a( g)$ is sufficiently regular. Generally asymptotic expansions have zero radius of convergence because
their coefficients grow {\it factorially}, \eqref{eq:ASBR6} being an example. 
\begin{defn}
An asymptotic expansion $\widetilde{f}$ of a function $f$ is said to be of Gevrey-$p$ if
\be\label{eq:ASBR12}
\widetilde f(g) = \sum_{n=0}^\infty c_n g^n, \qquad \text{with} \;\; |c_n| \leq K_1 K_2^n (n!)^p \,,
\ee
for some constants $K_1$ and $K_2$.
\end{defn}
The condition can also be expressed in terms of the finite function $f(g)$. We say that a function $f(g)$ admits a Gevrey-$p$ asymptotic expansion in a region $D$ if
\be\label{eq:ASBR13}
f(g) = \sum_{k=0}^{N-1} c_n g^n + R_N(g)\,, \quad \text{where} \quad |R_N(g)| \leq K_1 K_2^N (N!)^p  |g|^N\,,
\ee
for any $N$ and any $g\in D$. The perturbative loopwise expansions we encounter in QFT are generally Gevrey-1, so we will consider from now on this case only.

Asymptotic series are useful because they approximate the exact result for small values of $g$. The accuracy depends of course on $g$, but also on the behaviour 
of the series coefficients $c_n$ for $n\gg 1$. Contrary to convergent series, where the more terms are added in the series and the more accurate is the result, in asymptotic series there is an optimal number of terms one should keep, after which adding more terms results in worse and worse accuracy. This is called optimal truncation.
\begin{exr}
Suppose that for $n\gg 1$
\be
c_n \sim n! a^n \,. 
\label{eq:ASBR14}
\ee
For $a g>0$, show that the best accuracy for $f(g)$ is obtained when we keep in the sum
\be\label{eq:ASBR15}
 N_{{\rm Best}} = \left[\frac{1}{a  g} \right] \,,
 \ee
 where $[]$ indicates the integer part. Show also that the associated error, estimated as given by the last term not included in the sum, is proportional to
 \be\label{eq:ASBR16}
\Delta_{N_{\text{Best}}} \sim e^{-\frac{1}{a g }}\,.
\ee
\end{exr}
For small $g$ and $a\approx {\cal O}(1)$, \eqref{eq:ASBR16} shows that asymptotic series can be very precise, explaining e.g. the success of QED. 
But no matter how many terms we compute in perturbation theory, asymptotic series fail to reproduce the exact function by (at best) exponentially suppressed terms.
This is consistent with the intrinsic ambiguity related to asymptotic series shown in (\ref{eq:ASBR11}).
If the coupling $g$ is not that small, optimal truncation might not be sufficient and we should try to resum the series by resummation methods for divergent series.
We will in particular consider Borel resummations. 

\subsection{Basic notions of Borel resummations}
\label{subsec:basic}

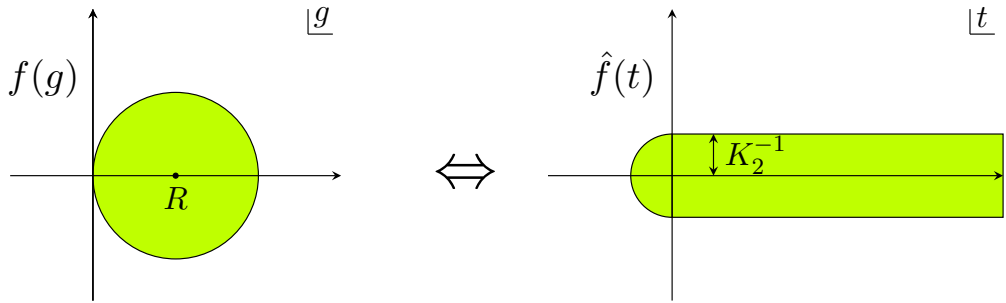
\begin{figure}[t!]
    \centering
    \raisebox{-3.em}
 {\scalebox{1.1}{
    \begin{tikzpicture}

\draw [-stealth] (0,-3/2) - - (0,2);
   \filldraw[fill=lime, draw=black]  (1,0) circle (1.cm);
  \draw [-stealth] (-1,0) - - (3,0);
    \filldraw[fill=black, draw=black]  (1,0) circle (0.03 cm);
\draw (0,-3/2) - - (0,2);
\draw (2.6,1.7) - - (2.6,2);
\draw (2.6,1.7) - - (2.9,1.7);
 \node[above] at (2.77,1.65) {\scalebox{1.}{$g$}};
  \node[above] at (-.6,.8) {\scalebox{1.2}{$f(g)$}};
 \node[below] at (1,0) {\scalebox{1.}{$R$}};
\node at (4.5,0) {\scalebox{2.}{$\Leftrightarrow$}};
\filldraw[fill=lime, draw=black] (7,1/2) arc (90:270:1/2);
\filldraw[fill=lime, draw=black] (7,-1/2) rectangle (11,1/2);
\draw [-stealth] (0,-3/2) - - (0,2);
\draw [-stealth] (5.5,0) - - (11,0);
\draw [-stealth] (7,-3/2) - - (7,2);
\draw (10.6,1.7) - - (10.6,2);
\draw (10.6,1.7) - - (10.9,1.7);
 \node[above] at (10.75,1.65) {\scalebox{1.}{$t$}};
 \draw [stealth-stealth] (7.5,1/2) - - (7.5,0);
 \node[right] at (7.5,1/4) {$K_2^{-1}$};
  \node[above] at (6.4,.8) {\scalebox{1.2}{$\hat f(t)$}};
\end{tikzpicture}}}
    \caption{Nevanlinna theorem: if a function $f(g)$ is analytic in the open green disc (left) and admits a Gevrey-1 expansion there, then 
    its Borel transform $\hat f$ is analytic in the green strip-like region (right), it is integrable and its inverse Laplace transform reproduces $f(g)$.
    The converse also applies.}
   \label{fig:Nevanlinna}
\end{figure}

Borel summation is a summation method for divergent series. Given $f(g)$ and its formal series expansion $\widetilde{f}(g)$,  
we define the Borel function $\hat f(t)$ as the function obtained by dividing the original series by a factorially growing factor:
\begin{equation}\label{eq:basic1}
\hat f(t)\equiv \sum_{n=0}^\infty \frac{c_n}{n!}t^n\,.
\end{equation}
If $\widetilde{f}$ is Gevrey-1, then by construction $\hat f$ is analytic in a disc around the origin and the series in \eqref{eq:basic1} has a non-zero radius of convergence 
where it defines the analytic function $\hat f(t)$. We denote Borel transforms of a function $f$ by a hat and by $t$ the argument of $\hat f$, also denoted
as the ``Borel plane".  If $\hat f(t)$ can be analytically continued over the $t$-plane beyond the disc where \eqref{eq:basic1} converges and is free of singularities in the positive real axis, $t>0$, then the integral
\begin{equation} 
f_{B}( g)=s(\widetilde f) \equiv \int_0^\infty dt\, e^{-t} \hat f(g t)\,,
\label{eq:basic2}
\end{equation}
if convergent, defines a function of $g$ that is said to be the Borel resummation of the asymptotic series $\widetilde{f}$.
In \eqref{eq:basic2} we have defined the Borel resummation operator $s$ acting on asymptotic series.
The function $f_B(g)$, by construction, has the same asymptotic series as the original function $f(g)$.
Indeed, we have
\be
f_B(g)=\int_0^\infty \! dt \, e^{-t}  \,\hat f(g t)\stackrel{?}{=}   \sum_{n=0}^\infty \frac{c_n}{n!}  g^n \int_0^\infty \! dt \,t^{n} e^{-t} = \sum_{n=0}^\infty c_n  g^n\,,
\label{eq:basic3}
\ee
and hence
\be
\widetilde f_B(g) = \widetilde f(g)\,,
\label{eq:basic4}
\ee
as formal power series.  Once again, the order of sum and integration in \eqref{eq:basic3} cannot be inverted because the Borel series expansion has generally a finite radius of convergence while the integral is taken over the whole positive $t$ axis. If we erroneously interchange the two operations, from a finite function $f_B$ we get its asymptotic divergent series, as occurred in \eqref{eq:ASBR5}. Given that \eqref{eq:basic4} holds, the key question is now: is $f_B=f$? 
In general this will not be the case. If, as we have shown, different functions can admit the same asymptotic series, it is clear that manipulating the latter cannot be enough to uniquely fix $f( g)$. A theorem, however, guarantees for us the necessary and sufficient conditions for Borel summability \cite{Nevanlinna,Sokal}:
\begin{thm}
{\bf (Nevanlinna's theorem \cite{Nevanlinna})} Let $f(g)$ be analytic in a open disc $D_R$ of radius $R$ whose center is at $g=R$ in the positive real axis (see figure \ref{fig:Nevanlinna}) and have there a Gevrey-1 asymptotic expansion. Then $\hat f(t)$ converges for $|t| < 1/K_2$ and has an analytic continuation with
\be\label{eq:basic5}
\hat f(t) \leq C e^{\frac{|t|}{R}} \,,
\ee
in a strip-like region which includes the whole positive real axis (see figure \ref{fig:Nevanlinna}), and
\be\label{eq:basic6}
f_B(g) = f(g)
\ee
in the disc $D_R$. The converse also applies.
\end{thm}
For a sketch of the proof see \cite{Sokal}. When $f=f_B$ we say that the asymptotic series $\widetilde{f}$ is Borel resummable to the exact result. In general, it is not easy to establish the analyticity conditions on $f$ to use the theorem, since in most cases these are unknown.\footnote{ When the asymptotic series arises from an integral, we can establish Borel summability using steepest descent arguments, bypassing Nevanlinna's theorem, see e.g. \cite{power}.} 

\begin{exr}
Consider the function $e^{-g^{-\alpha}}$ with $\alpha<1$.  Show that this function, analytic in $D_R$ for any $R$, 
does not satisfy \eqref{eq:ASBR12} in $D_R$.
\end{exr}

We can get some intuition on Borel functions by working out the Borel transform of an asymptotic series with coefficients exactly equal to the asymptotic behaviour (\ref{eq:ASBR14}), namely $c_n = n!a^n$. 
The Borel series in this case simply gives  
\be\label{eq:basic7}
{\hat f}(t) =\frac{1}{1-a t}\,.
\ee
Due to the simple pole at $t =1/a$, the radius of convergence of the Borel series is $R=1/|a|$, but the function \eqref{eq:basic7} can be analytically continued over the whole $t$-plane. Borel summability depends on the sign of $a$. If $a>0$  (same sign series) the singularity is on the positive real $t$ axis, the integral (\ref{eq:basic2}) is divergent and the series is not Borel resummable. If $a<0$ (alternating series), the singularity is over the negative real $t$ axis, the function $\hat f$ satisfies Nevanlinna's conditions with $K_2 = a$ and the integral (\ref{eq:basic2}) is finite. In the latter case the series is Borel resummable and the Borel resummed series is given by $f_B(g)$. 

More in general, (\ref{eq:ASBR14}) gives only the asymptotic form of the coefficients of the series, while the precise form of the latter might be unavailable.
When the exact Borel function ${\hat f}(t)$ is not known (the typical case)  some information on its analytic structure can still be deduced, because
 the large order behaviour of the asymptotic series determines the position of the singularity {\it closest} to the origin. 
If $a>0$ the series is not Borel resummable. On the other hand, if $a<0$ the series is not guaranteed to be Borel resummable, because 
further singularities on the positive real axis might occur, depending on the next to leading large order behaviour of the series coefficients. 

A generalization of the Borel transform (sometimes denoted Borel-Le Roy transformation), is obtained by defining
\be\label{eq:basic8}
{\hat f}_b(g) \equiv \sum_{n=0}^\infty \frac{c_n}{\Gamma(n+1+b)} g^n\,,
\ee
where $b$ is an arbitrary real parameter. The inverse Le Roy - Borel transform $f_B(g)$ reads
\be\label{eq:basic9}
f_B(g)=\int_0^\infty \! dt \,t^b e^{-t}  \,{\hat f}_b(g t)\,,
\ee
and is independent of $b$, as it should be. Borel-Le Roy functions with different $b$ can be related
analytically as follows (see e.g. \cite{power}):
\be \begin{split}\label{eq:basic10}
{\hat f}_b(t)&=t^{-b} \partial_t^{n}\left[t^{b+n} {\hat f}_{b+n}(t)\right] \,, \qquad n\in \mathbb{N}^+ \\
{\hat f}_{b+\alpha}(t)&=\frac{t^{-b-\alpha}}{\Gamma(\alpha)}\int_0^t dt'\,\frac{(t')^b\,{\hat f}_{b}(t')}{(t-t')^{1-\alpha}}\,, \qquad 0<\alpha<1\,.
\end{split} \ee
Note that the position of the singularities of two Borel-Le Roy transforms is the same, which implies 
that Borel summability does not depend on $b$, though the nature of the singularities in general changes. 

Before the advent of resurgence, Borel summability was considered essential. As we mentioned, proving it is not an easy task
because it requires to have control of the analyticity properties in the coupling of exact quantities in QFT. As a matter of fact, most
QFTs, including interesting ones such as gauge theories in $4d$, have loopwise expansions which are not Borel resummable.
Among the few exceptions where Borel summability has been established, 
bosonic theories with quartic interactions in $d<4$ dimensions ($\phi^4_d$ theories) are a notable class. 
More specifically, in \cite{Graffi:1970erh} it was found that the asymptotic series of the energy eigenvalues $\widetilde E_n(\hbar)$ of the $\phi_1^4$ model (the quartic quantum mechanical anaharmonic oscillator) is Borel resummable. In  QFT ($d=2$ and $d=3$) the existence of a non-perturbative definition of the theory is generally the first important point to be established, Borel summability (if any) being a by-product. This proof has been given in \cite{Eckmann,Magnen} for the Euclidean $\phi^4_d$ theories ($d=2,3$) in a specific renormalization scheme, in which $2n$-point smeared Schwinger functions of local operators have been shown to satisfy Nevanlinna's conditions in an infinitesimal disc in the coupling constant complex plane.\footnote{See \cite{Serone:2018gjo,Serone:2019szm} for an heuristic, but much simpler and more general, argument for the Borel summability of such theories using steepest descent arguments.}

\subsection{Large orders in perturbation theory: instantons and complex instantons}
\label{subsec:LOPT}

Given the difficulties of establishing rigorously the Borel summmability of asymptotic expansions in QFT,
early works  were mostly interested in determining their large order behaviour \cite{Bender:1969si,Lipatov:1976ny,Brezin:1976wa,Brezin:1976vw} (see  \cite{guillou1990large} for a collection of early papers), namely the coefficient $a$ in \eqref{eq:ASBR14}, and possibly other coefficients governing  sub-leading corrections.
The motivation was two-fold:
\begin{itemize}
\item The sign of $a$ can be taken as a first indication of the Borel summability of the series.
\item When $a<0$ or we know the theory is Borel resummable (like in $\phi^4_d$ theories), the knowledge of the coefficients entering the large order behaviour,
especially $a$, is useful to numerically reconstruct the Borel function from its first few known perturbative coefficients.
In particular, under some assumptions, the knowledge of $a$ allows to conformally map the original asymptotic series to a convergent one.
\end{itemize}
The large order behaviour in QFT is determined by certain saddle points in a complexification of the theory \cite{Lipatov:1976ny}.
We show the key idea for the ordinary integral \eqref{eq:ASBR4}. The coefficients of its perturbative expansion are given in \eqref{eq:ASBR6}.
Lipatov's trick consists in writing $c_n$ as a contour integral
\be
c_n = \frac{1}{2i\pi} \oint_{{\cal C}_0} \frac{dg}{g^{n+1}}\widetilde I(g) \sim \frac{1}{2i\pi} \oint_{{\cal C}_0} \frac{dg}{g^{n+1}} \int_{-\infty}^\infty \! dx\, e^{-\frac{x^2}{2}-\frac{gx^4}{4}}\,,
\label{eq:LOPT1}
\ee
where ${\cal C}_0$ is a small contour around the origin $g=0$ and we have naively replaced the formal power series with the actual function in \eqref{eq:LOPT1}.\footnote{We will later mention more proper ways to get the large order behaviour. For now it suffices to say that this naive method works.} We evaluate the double integral in \eqref{eq:LOPT1} by a saddle-point approximation. We change variables $x = g^{-1/2} y$ and then extend both integrals in the complex plane. In this way we get, omitting numerical factors, 
\be\label{eq:LOPT2}
c_n \approx  \oint_{{\cal C}_0} \! \! dg \! \int_{\cal C} \! dz\, e^{-S(z,g)}\,, \qquad
S(z,g) = \frac{1}{g} \Big(\frac{z^2}{2}+\frac{z^4}{4}\Big) + \Big(n+\frac 32\Big) \log g \,,
\ee
where ${\cal C}$ is a deformation of the path running over the real axis. 
For $n\gg 1$ and $g\ll 1$ a saddle-point approximation is valid for both integrals at the same time.
We get three extrema for $S$. One at $g=\infty$ and $z=0$, which is singular and should be neglected, and two  at
\be\label{eq:LOPT3}
z_0=\pm i \,, \qquad \qquad g_0 = -\frac{1}{4\big(n+\frac 32\big)}\,.
\ee
Plugging the two (equivalent) non-trivial saddles in $S(z,g)$ we get the leading order approximation of $c_n$:
\be\label{eq:LOPT4}
c_n\approx n! (-4)^n \ldots 
\ee
In order to reliably compute $\ldots$ we should also consider the quadratic fluctuations around the saddle. 
When these are taken into account, the large order behaviour in \eqref{eq:ASBR6} is precisely reproduced.
We leave to the reader to perform this straightforward check.

We have then established a connection between the large order behaviour of an asymptotic series and saddle points. 
For ordinary integrals this connection can be generalized and made more rigorous \cite{Berry2}.
Let
\be
I^{(0)}(g) = \frac{1}{\sqrt{g}} \int_{{\cal C}_0} \! dz\, e^{-\frac{f(z)}{g}} \sim \sum_{p=0}^\infty c_p^{(0)} g^p\,,
\label{eq:LOPT5}
\ee
where $f(z)$ is an analytic function and ${\cal C}_0$ is a regular path of steepest-descent (also called Lefschetz thimble) passing through the saddle point $z_0$ of $f$. 
For simplicity let $f$ be an entire function and $z_0$ be an isolated and non-degenerate saddle point. Given $z_0$, we can define saddles adjacent to it as follows.
Consider the integral (\ref{eq:LOPT5}) as a function of $g=|g|\exp(i\theta)$. The path of steepest descent moves in the complex $z$-plane as $\theta$ is varied. 
Saddles which are crossed by the steepest descent path as $\theta$ is varied are called  ``adjacent" to $z_0$. 
Among the adjacent saddles, we denote by $z_{\sigma_0}$ the leading adjacent saddle as the one with the smallest value of $|f(z_{\sigma_0})-f(z_0)|$.
Then, the following relation holds \cite{Berry2}:
\be
c_p^{(0)}  \underset{p\to\infty}{\approx} \sum_{z_{\sigma_0}} \frac{\Gamma(p)}{\big(f(z_{\sigma_0})-f(z_0)\big)^p} c_0^{(\sigma_0)}\,,
\label{eq:LOPT6}
\ee
where
\be\label{eq:LOPT7}
c_0^{(\sigma_0)} = \frac{1}{\sqrt{2\pi |f^{\prime\prime}(z_{\sigma_0})|}}
\ee
is the leading (one-loop) coefficient term around the adjacent saddle(s) and the sum is present in case we have more than one leading adjacent saddle. 
It is immediate to recover \eqref{eq:ASBR6} applying \eqref{eq:LOPT6}
to the integral \eqref{eq:ASBR4}. We see that the large order behaviour of a perturbative expansion around a saddle point is given by the leading terms of ``nearby" saddles, 
a typical manifestation of resurgence, as we will repeatedly see in these lectures.

The idea can be applied to quantum mechanics and quantum field theory as well. For example, for the quartic anharmonic oscillator 
with potential $V(q) = q^2/2+q^4/4$, the large order behaviour of the ground state energy $E_0\sim \sum_p c_p \hbar^p $ is given by 
\be
c_p\underset{p\to\infty}{\propto}  \oint_{{\cal C}_0} d\hbar \!\int\!{\cal D}q(t)\exp\Big(-p \log \hbar -\frac{1}{\hbar} \int_{-\infty}^\infty \! dt\, L(q)\Big)\,,\qquad
L(q) = \frac{\dot q^2}{2}+\frac{q^2}{2}+\frac{q^4}{4}\,,
\label{eq:LOPT8}
\ee
where ${\cal C}_0$ is a small contour around the origin $\hbar =0$, and ${\cal D}q(t)$ is the path integral measure for all possible complex paths. 
Extremizing with respect to $\hbar$ gives
\be\label{eq:LOPT9}
-\frac{p}{\hbar_0} + \frac{S_0}{\hbar_0^2} = 0\,,
\ee
where $S_0$ is the smallest non-trivial classical configuration with finite action $S_0$. There are no real, finite action instanton configurations for the quartic anharmonic oscillator, but we do have {\it complex} instanton configurations. In particular, it is easy to verify that the purely imaginary instanton 
\be\label{eq:LOPT10}
q(t) = \frac{i \sqrt{2}}{\cosh t}\,,
\ee
satisfies the equation of motion and has $S_0 = -4/3$. Plugging back in \eqref{eq:LOPT8} and proceeding as before, we get
\be\label{eq:LOPT11}
c_p \propto p! S_0^{-p}\,.
\ee
A more precise estimate can be performed by taking into account fluctuations around the instanton background. See \cite{Brezin:1976vw} for details.
Once again, we see that the large order behaviour of a perturbative expansion is given by other saddles, complex instantons in this case.
Like ordinary integrals, rigorous results for the large order behaviour of the perturbative expansion  can also be obtained in quantum mechanics using resurgence \cite{AKT1,reshyper,ddpham,dpham,AKT2}.

The method of \cite{Lipatov:1976ny} has been used also in QFT, though no rigorous derivation has been established there.\footnote{The key point is to understand which saddles should be considered, i.e. the analogues of the adjacent saddles discussed above for ordinary integrals.} 
The large order behaviour of quartic scalar theories in $d=2,3$ dimensions has been found to be given by purely imaginary instantons and is sign alternating, compatibly
with the Borel summability of the perturbative expansion in these theories, as discussed at the end of section \ref{subsec:basic}.
Whenever a QFT admits real instantons, one has $S_0>0$, a same sign series, and Borel summability is lost. This is of course expected, since instanton contributions should be added to the perturbative result. Each instanton saddle gives rise to its asymptotic coupling expansion and we naturally get trans-series.

\subsection{Applications in critical phenomena}

Resummation methods have been applied successfully for decades  in the study of second order phase transitions in critical phenomena.
In this context, the two main approaches are:
\begin{itemize}
\item{$\epsilon$-expansion \cite{Wilson:1971dc}: the large order behaviour of the $\epsilon$-expansion for quartic scalar models was computed in \cite{Brezin:1976vw}
and found to be sign-alternating. Since then, assuming that the series is Borel resummable (still unproven) several critical exponents have been computed.
Starting from a truncated series expansion it is possible to numerically reconstruct an approximant for the Borel function associated to the observable, which is then 
computed by the Laplace  transformation \eqref{eq:basic2}.}
\item{Fixed dimension \cite{Parisi:1980gya}:  the space-time dimension is fixed to its physical value and we start from a massive disordered phase. The large order behaviour of the coupling expansion for quartic scalar models was computed in \cite{Lipatov:1976ny}, refined in \cite{Brezin:1976vw}, and found to be
sign-alternating. The perturbative series has been proved to be Borel resummable \cite{Eckmann,Magnen}.
The critical point is found by setting to zero a Borel resummed $\beta$-function for the quartic scalar coupling, from which critical exponents are computed 
by a Borel resummation of the coupling expansion evaluated at the critical value. The Borel resummations are performed numerically 
using algorithms similar to those used for the $\epsilon$-expansion.}
\end{itemize} 
The accuracy obtained over the years with the two methods is similar. Using $<10$ coefficient terms, an accuracy of order $1\%-1\permil$ is obtained, see e.g. \cite{Kompaniets:2017yct} for an account for the  $3d$ $O(N)$ vector models in the $\epsilon$-expansion.\footnote{The accuracy is estimated by comparing the results with more precise techniques (when available), such as the conformal bootstrap \cite{Poland:2018epd}.}
Both methods have pros and cons. The $\epsilon$-expansion allows for an analytic understanding of the fixed points for $\epsilon\ll1$ and requires the computation of massless loops, which can be handled analytically (see \cite{Schnetz:2022nsc} for a recent account in the $\phi^4$ theory). The fixed dimension expansion requires instead the harder computation of loops involving massive particles, which is typically handled numerically at high loops (see \cite{Sberveglieri:2023mzy} for recent progress in this direction).
The latter method is however more rigorous, since Borel summability is established.\footnote{Strictly speaking, Borel summability has not been established in 
the renormalization scheme of \cite{Parisi:1980gya} and used since then, but in others. See \cite{Sberveglieri:2019ccj} for details on this point.}
 It also allows us to study the massive phase, away from criticality \cite{Serone:2018gjo,Serone:2019szm}, and comparisons with other non-perturbative methods are possible.
 
The knowledge of the coefficient $a<0$  governing the large order behaviour of the asymptotic series in \eqref{eq:ASBR14} is important 
in setting up algorithms to numerically reconstruct the Borel function from a truncated series. In particular, under the assumption that all Borel singularities sit on the negative real axis, we can perform a conformal transformation $t=t(u)$ to map the Borel plane $t$ to the unit disc $|u|\leq 1$. The transformed Borel function $\widetilde B(u) = B(t(u))$ converges for $|u|<1$ and as a result the mapping converts the original asymptotic series into a convergent one. This conformal mapping technique is one of the main and efficient methods used to resum asymptotic series, though it is useful to recall that the assumption about the location of the Borel singularities is yet unproven. We will not discuss further details of the conformal mapping or of other numerical methods.

\subsection{Obstruction to Borel summability: renormalons}
\label{subsec:reno}

\begin{figure}[t!]
    \centering
    \raisebox{-3.em}
 {\scalebox{1.6}{
    \begin{tikzpicture}
\draw[rotate=90]  (1,0) ellipse (10pt and 5pt); 
\draw[rotate=60]  (1.55,0.65) ellipse (10pt and 5pt); 
\draw[rotate=30]  (1.7,1.48) ellipse (10pt and 5pt); 
\draw[rotate=0]  (1.4,2.3) ellipse (10pt and 5pt); 
\draw[rotate=0]  (3,1) ellipse (5pt and 10pt); 
\draw[-stealth] (-0.5,0.64) -- (1.5,0.64);   
\draw (1.5,0.64) -- (3.5,0.64);   
     \filldraw[fill=black, draw=black]  (0.,0.64) circle (0.03 cm);  
         \filldraw[fill=black, draw=black]  (3.,0.64) circle (0.03 cm);  
\draw[dashed] (3,1.5) arc (10:90:10mm);
        \filldraw[fill=black, draw=black]  (0.03,1.35) circle (0.03 cm);  
         \filldraw[fill=black, draw=black]  (0.42,1.95) circle (0.03 cm);    
         \filldraw[fill=black, draw=black]  (1.06,2.27) circle (0.03 cm);  
                  \node[below] at (-0.35, 0.55) {\scalebox{.6}{$p$}};
                      \node[below] at (1.35, 0.57) {\scalebox{.6}{$p+k$}};
                            \node[below] at (3.35, 0.55) {\scalebox{.6}{$p$}};
            
  \end{tikzpicture}}}
    \caption{Renormalon $n+1$-loop diagram contributing to the two-point function of $\phi$ in the $\phi^4$ theory. The diagram consists in a chain of $n$ small bubbles and a large 
loop over the momentum $k$.}
   \label{fig:Renormalon}
\end{figure}
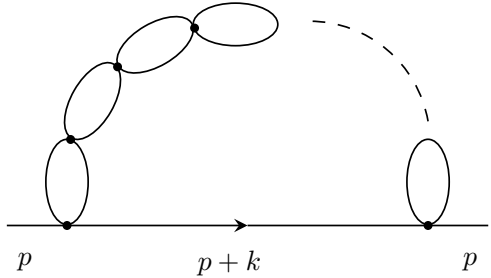

Instantons govern the large order behaviour of an asymptotic series whenever the factorial growth of the perturbative coefficients is given by the multiplicity of Feynman diagrams. Namely, barring
cancellations among diagrams at a given loop level, we have ${\cal O}(1)$ contributions from ${\cal O}(n!)$ Feynman diagrams.
Another source of factorial growth exists, which gives rise to  ${\cal O}(n!)$ contributions from ${\cal O}(1)$ Feynman diagrams.
The associated singularity in the Borel plane is called a renormalon singularity. Renormalons arise in theories with classically, but not exactly, marginal couplings. 
The name ``renormalon" is due to the their link with the RG flow of the couplings which, in turn, can be determined from the UV renormalization of the theory.
In fact, they typically arise from $n$-loop diagrams given by a chain of the one-loop diagrams entering the determination of the leading order $\beta$-function for a given coupling.
See fig.\ref{fig:Renormalon} for an example in the $\phi^4$ theory in $d=4$ dimensions. The basic mechanism giving the renormalon singularity is universal.
After renormalization, one-loop diagrams entering in the renormalon chain depends only on the momentum of the large loop (indicated with $k$ in figure \ref{fig:Renormalon})
and is proportional to $\beta_0 \log (k^2/\mu^2)$, where $\beta_0$ is the one-loop coefficient of the $\beta$-function and $\mu$ is the RG scale. The $n$-loop diagrams,
together with all counterterms insertions to make them UV finite, are proportional to\footnote{We neglect scheme dependent threshold terms,  irrelevant 
in our considerations.}
\be
{\cal G}_n \sim \int_0^\infty \!\! dk \, F(k) \Big(\beta_0 \log \frac{k}{\mu}\Big)^n\,,
\label{eq:reno1}
\ee
where $k$ denotes the radial component of the momentum $k^\mu$ and $F(k)$ is a model-dependent function which we do not need to specify. The integral \eqref{eq:reno1} can have two sources of factorial growth, in the IR or in the UV. Suppose that 
\be
\begin{split}
F(k) &\underset{k\to0 }{\sim} k^{a_{{\rm IR}}-1}\,, \qquad \quad\;  a_{{\rm IR}}>0\,, \\
F(k) &\underset{k\to\infty}{\sim} k^{-a_{{\rm UV}}-1}\,, \qquad a_{{\rm UV}}>0\,.
\end{split}
\label{eq:reno2}
\ee

We can split $\int_0^\infty = \int_0^1+\int_1^\infty$ and, neglecting subleading contributions, we replace in the two integrals the behaviour \eqref{eq:reno2} to get
\be
\begin{split}
\int_0^1\!\! dk \,   k^{a_{{\rm IR}}-1}  \Big(\beta_0 \log \frac{k}{\mu}\Big)^n & = n! \Big(\frac{-\beta_0}{a_{{\rm IR}}}\Big)^n a_{{\rm IR}}\,, \\
\int_1^\infty\!\! dk \,  k^{-a_{{\rm UV}}-1}  \Big(\beta_0 \log \frac{k}{\mu}\Big)^n & = n! \Big(\frac{\beta_0}{a_{{\rm UV}}}\Big)^n a_{{\rm UV}}^{-1}\,.
\end{split}
\label{eq:reno3}
\ee
A few comments are in order.
\begin{itemize}
\item The $n$-loop diagrams associated to ${\cal G}_n$ lead to two singularities in the Borel plane, at 
\be\label{eq:reno4}
\begin{split}
t & = - \frac{a_{{\rm IR}}}{\beta_0} \,, \qquad \text{(IR Renormalon)} \\
t & =  \frac{a_{{\rm UV}}}{\beta_0} \,, \qquad \text{(UV Renormalon)} \,.
\end{split}
\ee
\item Independently of the sign of $\beta_0$, Borel summability is lost.
For $\beta_0>0$ this is due to the UV renormalons and is related to the fact that the theory is non-perturbatively
ill-defined. For $\beta_0<0$ this is due to the IR renormalons and is related to the fact that perturbation theory breaks down at low energies.
\item Renormalons reflect  the mathematical inconsistency of a perturbative expansion in a parameter which is dimensionless only at the classical level.
\item If a theory is perturbatively gapped in the IR, e.g. we have a mass term in the $\phi^4$ theory above, the integral in \eqref{eq:reno1} has a IR cut-off at the mass scale and the IR renormalon
disappears. In contrast, UV renormalons are always present.
\item Renormalons are often the leading source of factorial growth of the perturbative expansion and dominate over singularities induced by instantons (or instantons anti-instantons) configurations.
\end{itemize}
Renormalons were first discovered in the $2d$ Gross-Neveu model in the original paper \cite{gross-neveu}, they were found in QED (in the study of $g-2$) in \cite{Lautrup:1977hs} and discussed by 't Hooft (who coined the term renormalon) in QCD in \cite{tHooft:1977xjm}. See \cite{beneke} and references therein for further details.

The existence of renormalons in $4d$ gauge theories is hard to prove, because at order $n$ many other diagrams contribute. Cancellation among diagrams is possible, or the presence of other factorially growing diagrams which are not of the renormalon-type.  The chain of bubble diagrams entering renormalons can however be shown to be dominant in the special case of large $N$ in vector models (not to be confused with 't Hooft large $N$ limit). Since a few years ago, the existence of renormalons was rigorously established only for some $2d$ vector models at large $N$ such as the Gross-Neveu \cite{gross-neveu} and the non-linear sigma models \cite{David:1982qv,David:1983gz,Novikov:1984ac}. 

In contrast to the factorial growth due to the diagram multiplicity, renormalons do not have a semi-classical interpretation, or at least this has not been found so far.
On the other hand, whenever the observable admits an OPE expansion, renormalons have been argued to be related to power corrections \cite{Shifman:1978bx}.
Let us schematically describe the origin of such correspondence in the context of a UV free $4d$ non-abelian gauge theory which confines in the IR.
Let $\Phi$ be a generic gauge-invariant composite local operator.  Using the OPE, we can write its two-point function when $x\rightarrow 0$ as
\be
\langle \Phi(x) \Phi(0) \rangle \sim \sum_n c_n(\mu x) \langle O_n(0) \rangle x^{\Delta_{O_n}}\,,
\label{eq:reno5}
\ee
where $\mu$ is the RG scale, the sum runs over all gauge invariant operators $O_n$  with a non-trivial one-point function on the vacuum (condensates) and $\sim$ indicates that the
sum in \eqref{eq:reno5} might generally be asymptotic. Let $\Delta_{O_n}$ be the scaling dimensions of $O_n$ in the free UV  CFT. 
By dimensional analysis, we  have
\be\label{eq:reno6}
\langle O_n \rangle \sim \Lambda^{\Delta_{O_n}}\,,
\ee
where $\Lambda$ is the dynamically generated scale of the theory, 
\be\label{eq:reno6a}
 \Lambda \approx \mu e^{\frac{1}{g(\mu)\beta_0}}\,.
\ee
Each condensate gives rise to a non-perturbative contribution to the correlator,
whereas the coefficient functions $c_n(\mu x)$ encode the perturbative contributions around the condensate and are given in general by an asymptotic series in the coupling.\footnote{The splitting 
in the OPE \eqref{eq:reno5} between $c_n$ and the condensate is not uniquely defined and in particular depends on the renormalization scheme. See e.g. \cite{beneke} for further details
in a renormalon perspective.} The coefficient associated to the identity operator exchange corresponds to the ordinary perturbative expansion. 
If the latter features IR renormalon singularities as in \eqref{eq:reno4}, the series is not Borel resummable because of the singularity in the integral \eqref{eq:basic2}.
Alternatively we can deform the contour of integration to avoid the singularity, in which case however we do not get a well-defined result. 
A way to estimate the size of the ambiguity is to assume that the singularity is a simple pole and compute its residue:
\be\label{eq:reno7}
\frac{1}{2i \pi} \oint_{t_0} \! dt \frac{e^{-\frac{t}{g}}}{t-t_0} = e^{-\frac{t_0}{g}}\,, \qquad t_0 = \frac{-a_{\rm IR}}{\beta_0}\,.
\ee
The non-perturbative contribution \eqref{eq:reno7} nicely matches the condensate \eqref{eq:reno6}, provided we identify
\be\label{eq:reno8}
a_{\rm IR} = \Delta_{O_{\rm min}}\,,
\ee
where  $\Delta_{O_{\rm min}}$ is the smallest scaling dimension (besides the identity) of the operators $O_n$ appearing in  the OPE \eqref{eq:reno5}.
Higher order operators are expected to give rise to further singularities over the positive real axis at $t_n = - \Delta_{O_n}/\beta_0$.
Renormalon singularities can be then associated to (non-perturbative) power corrections in correlation functions admitting an OPE expansion.
Sharp results on this correspondence are limited to 2d models at large $N$ \cite{David:1982qv,David:1983gz,Novikov:1984ac,Beneke:1998eq}. In fact, most of the interest on renormalons in 2d large $N$ models
was motivated by verifying this ``renormalon-OPE"  correspondence. The correspondence is assumed to hold in $4d$ QCD and it is at the base of
application of renormalon physics in QCD phenomenology. 

\section{Basics of resurgence}
\label{sec:resurgence}

We have seen in section \ref{sec:ASBR} that  an infinite number of complex functions can lead to the same asymptotic series expansion $\widetilde f(z)$. 
Due to the Stokes' phenomena, it can happen that the Borel resummation of $\widetilde f$ gives a function in a region of the complex plane, but some other function
in some other region. These regions are delimited by Stokes lines. 
Resurgence is essentially a systematic way to take into account of the Stokes phenomenon and give a meaning to otherwise non-Borel resummable asymptotic series.
Its foundations  have been laid by J. Ecalle in his seminal works \cite{ecalle}. Unfortunately, some proofs are still missing and resurgence is not a fully rigorous
mathematical subject yet. 

It is useful to generalize the integral \eqref{eq:basic2} along an arbitrary ray with angle $\theta$ in the complex $t$-plane by defining
\be\label{eq:resurgence1}
s_{\theta}(\widetilde f) = \frac{1}{g}\int_0^{e^{i \theta}\infty}\! \! dt\, e^{-\frac{t}{g}} \hat f(t)\,, 
\ee
where we require 
\be\label{eq:resurgence2}
{\rm Re}\,\Big(\frac{e^{i\theta}}{g}\Big)>0
\ee
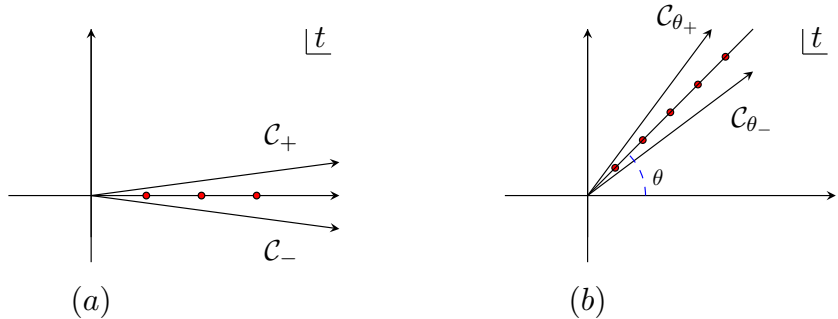
\begin{figure}[t!]
    \centering
    \raisebox{-3.em}
 {\scalebox{1.1}{
    \begin{tikzpicture}

\draw [-stealth] (0.,-.5) - - (0,2);
   \draw [-stealth] (-1,0) - - (3,0);
  \draw (0,-.8) - - (0,2);
\draw (2.6,1.7) - - (2.6,2);
\draw (2.6,1.7) - - (2.9,1.7);
  \filldraw[fill=red, draw=black]  (2/3,0) circle (0.04cm);
  \filldraw[fill=red, draw=black]  (4/3,0) circle (0.04cm);
  \filldraw[fill=red, draw=black]  (2,0) circle (0.04cm);
\draw[-stealth] (0.,0.) - - (3.,0.4);
\draw[-stealth] (0.,0.) - - (3.,-0.4);
 \node[above] at (2.3,.4) {\scalebox{.9}{${\cal C}_+$}};
  \node[below] at (2.3,-.4) {\scalebox{.9}{${\cal C}_-$}};
\node[above] at (2.77,1.65) {\scalebox{1.}{$t$}};
\node[below] at (0.,-.95) {\scalebox{1.}{$(a)$}};

\draw [-stealth] (6.,-.5) - - (6.,2);
   \draw [-stealth] (5,0) - - (9,0);
  \draw (6,-.8) - - (6,2);
\draw (8.6,1.7) - - (8.6,2);
\draw (8.6,1.7) - - (8.9,1.7);
  \filldraw[fill=red, draw=black]  (19/3,1/3) circle (0.04cm);
  \filldraw[fill=red, draw=black]  (20/3,2/3) circle (0.04cm);
  \filldraw[fill=red, draw=black]  (7,1) circle (0.04cm);
    \filldraw[fill=red, draw=black]  (22/3,4/3) circle (0.04cm);
  \filldraw[fill=red, draw=black]  (23/3,5/3) circle (0.04cm);

\draw[-stealth] (6.,0.) - - (7.5,2.);
\draw[-stealth] (6.,0.) - - (8.,1.49);
\draw (6.,0.) - - (8.,2.);
 \node[above] at (7.1,1.8) {\scalebox{.9}{${\cal C}_{\theta_+}$}};
  \node[below] at (8.,1.2) {\scalebox{.9}{${\cal C}_{\theta_-}$}};
\node[above] at (8.77,1.65) {\scalebox{1.}{$t$}};
\node[below] at (6.,-.95) {\scalebox{1.}{$(b)$}};
\draw[blue,dashed](6.7,0) arc (0:45:.7);
\node[below] at (6.85,.45) {\scalebox{0.7}{$\theta$}};
\end{tikzpicture}}}
    \caption{Contours associated to lateral Borel resummations: (a) around the real positive axis, (b) around an arbitrary ray at angle $\theta$ from the positive real axis.}
   \label{fig:lateral}
\end{figure} 
in order to have a convergent integral. Whenever an asymptotic series $\widetilde f(z)$ is not Borel resummable because of singularities occurring along the positive real axis in $\hat f(t)$, 
assuming that such singularities are not dense in $t$ and form a natural boundary, we can deform the original contour into contours ${\cal C}_\pm$ which go in the upper or lower plane, as in fig.\ref{fig:lateral} (a). The non-Borel summability is quantified by the difference of Borel resummation along ${\cal C}_\pm$: 
\be\label{eq:resurgence2a}
(s_+-s_-)(\widetilde f)\neq 0\,,
\ee
where $s_\pm \equiv s_{\pm \epsilon}$, with $\epsilon\ll 1$, define so called lateral Borel resummations. We define the Stokes automorphism $\mathfrak{S}$  as
\be\label{eq:resurgence3}
s_+ \equiv s_- \mathfrak{S} = s_- ({\rm Id} - {\rm Disc})\,, \qquad s_+-s_- = - s_- {\rm Disc}\,.
\ee
If the singularities occur along a ray with angle $\theta$, we can similarly define the Stokes automorphism $\mathfrak{S}_{\theta}$ along a ray as
\be\label{eq:resurgence4}
s_{\theta^+} = s_{\theta^-} \mathfrak{S}_{\theta}\,, \qquad s_{\theta^+}-s_{\theta^-} = - s_{\theta^-} {\rm Disc}_\theta\,, \qquad  \mathfrak{S}_{\theta=0}=  \mathfrak{S}\,.
\ee
The key idea underlying resurgence is to reconstruct the function $f$ by the knowledge of the singularities of $\widetilde f$ (which from now on we denote by $\widetilde f_0$). 
For example, let $a_n$ be the positions of the singularities (in general branch-cut singularities) along the positive real axis in the Borel plane with $a_{n}<a_{n+1}$, $n=1,2,\ldots$,
 as in panel $(a)$ of figure \ref{fig:lateral}. The ambiguity to $s(\widetilde f_0)$, quantified by \eqref{eq:resurgence2a}, is given by the closest singularity $a_1$ and is of order
$\exp(- a_1/g)$. A more careful investigation shows that $\exp(- a_1/g)$ multiplies in general an entire new formal power series $\widetilde f_1$. This series can also give rise
to ambiguities, which will be suppressed by a factor $\exp(- 2a_1/g)$, and again lead to the discovery of another formal power series $\widetilde f_2$, and so on.\footnote{In general, if the ambiguity of $s(\widetilde f_n)$ contains sub-leading exponentially suppressed terms, the situation is more complicated. The various formal power series mix in an entwined way and we can have
exponentially suppressed terms of the general form $\exp(- \sum_{k=1} p_k a_k/g)$, with $p_k\in \mathbb{N}$.
We focus here on the simple case where the ambiguities are determined by $a_1$ and are proportional to $\exp(- n a_1/g)$, $n\in \mathbb{N}$, since this corresponds to the case discussed in section \ref{sec:2d1N}.}
In general, the process requires the addition of an infinite number of formal power series $\widetilde f_n$.
An entire non-perturbative sector is unveiled by the Stokes phenomenon. We then replace the perturbative series expansion $\widetilde f_0$ of $f$ with a trans-series $\widetilde F$:
\be\label{eq:resurgence4b}
\widetilde F(g) = \sum_{n=0}^\infty \sigma_n e^{-\frac{a_n}{g}} \widetilde f_n(g)\,.
\ee
Although the formal series $\widetilde f_n$ might not be Borel resummable individually, the whole trans-series $\widetilde F$ could, thanks to a cancellation of the ambiguities.
Roughly speaking, resurgence is the theory underlying this reconstruction process. The coefficients $\sigma_n$ in \eqref{eq:resurgence4b} are denoted Stokes constants. 
Their determination in general is non-trivial. In the simple case we will discuss, we have $\sigma_n = \sigma^n$, and we will show how to fix $\sigma$.

A general framework where resurgence is useful is in solving differential equations starting from an ansatz in terms of asymptotic series. 
Whenever asymptotic series comes from a (path) integral, the Stokes discontinuity is related to a jump of steepest descent paths.
We start with a gentle introduction to resurgence by working out the simple and historically notable case of the Airy function in sections \ref{subsec:AiryDE} and \ref{subsec:AiryInt}.
After that, we go back to general considerations in section \ref{subsec:alien}.

\subsection{The Airy function as solution of a differential equation}

\label{subsec:AiryDE}

\begin{figure}[t!]
    \centering
    \raisebox{-3.em}
 {\scalebox{1.4}{
    \begin{tikzpicture}
\draw[snake=snake,segment length=5pt,red] (-1,3)-- (1,3); 

   \filldraw[fill=black, draw=black]  (1,3) circle (0.05 cm);

       \draw [stealth-] (2,3) - - (3,3);
            \draw (1,3) - - (2,3);
              \draw [-stealth]  (1.,3.) to (0.5,3.866);
         \draw (0.5,3.866) to (0,4.732);
      \draw [-stealth]  (1.,3.) to (0.5,2.134);
           \draw (0.5,2.134) to (0,1.268);
                  \node[below] at (2,2.1) {\scalebox{1.2}{I}}; 
                    \node[above] at (2,3.9) {\scalebox{1.2}{II}}; 
                         \node[above] at (-1.5,2.9) {\scalebox{1.2}{III}}; 

 \end{tikzpicture}}}
    \caption{The three wedges of the complex plane delimited by Stokes lines and their orientation for the Airy function. The branch-cut (red) is also reported.} 
	\label{fig:Airy}

\end{figure}
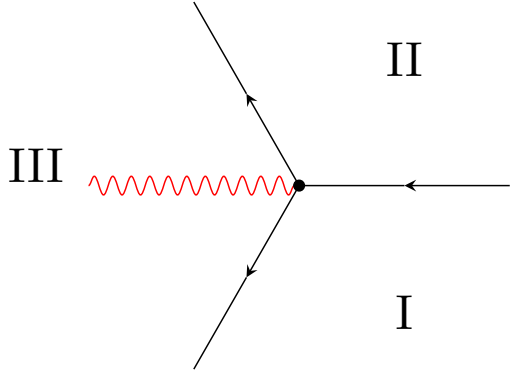

The Airy differential equation reads
\be
y^{\prime\prime}(x)  =  x y(x)\,,
\label{ODI1}
\ee
where $x\in \mathbb{R}$. We look for solutions obtained via an asymptotic expansion. In order to do that, we insert a dummy variable $\epsilon$ and consider the complexified
equation with $x\rightarrow z \in \mathbb{C}$:
\be
\epsilon^2 y^{\prime\prime}(z)  =  z y(z)\,.
\label{ODI2}
\ee
The leading terms of an expansion for $\epsilon\ll 1$ are given by
\be
y_\pm \approx z^{-\frac 14} e^{\pm \frac{2}{3\epsilon} z^{3/2}}\,.
\label{ODI3}
\ee
We then look for a solution in terms of a power series:
\be
\widetilde y_\pm = \frac{1}{2\sqrt{\pi}}z^{-\frac 14} e^{\pm \frac{2}{3\epsilon} z^{3/2}}  \widetilde \psi_\pm(z)\,, \qquad \widetilde \psi_\pm(z) = \sum_{n=0}^\infty  \alpha_n^{\pm}(z) \epsilon^n \,,
\label{ODI4}
\ee
where $\alpha_0^\pm(z)=1$ and we have conveniently normalized the series.  As we momentarily see, the series above is asymptotic and this explains the use of the tilde symbol.
Large $z$ is equivalent to small $\epsilon$ and the $z$-dependence of the $\alpha_n$ is easily found:
\be
\alpha_n^\pm(z) = c_n^\pm \,  z^{-\frac{3n}{2}}\,.
\label{ODI5}
\ee
Plugging the ansatz \eqref{ODI4} in \eqref{ODI2} allows us to fix all the coefficients $c_n$.
One gets 
\be
c_n^{\pm} =  \Big(\pm \frac 34 \Big)^n \frac{\Gamma\Big(n+\frac 16\Big) \Gamma\Big(n+\frac 56\Big)}{2\pi\Gamma(n+1)}\,.
\label{ODI6}
\ee
\begin{exr}
Derive \eqref{ODI6}.
\end{exr}
We compute the Borel transforms $\hat \psi_\pm$:
\be
\hat{\psi}_{\pm}(t) =  \sum_{n=0}^\infty \frac{c_n^\pm}{n!} (t z^{-3/2})^n = {}_2F_1\Big(\frac 16, \frac 56, 1; \pm \frac{3t}{4z^{\frac 32}}\Big)\,.
\label{ODI7}
\ee
The functions $\hat{\psi}_{\pm}$ are analytic in the cut complex $t$-plane, with a branch-cut singularity at $t_\pm  = \pm 4 z^{3/2}/3$.
For complex $z$ both $\hat{\psi}_{+}$ and $\hat{\psi}_{-}$ are Borel resummable. By taking the Laplace transform of $\hat \psi_\pm$ as in \eqref{eq:basic2}
we get two well-defined functions which correspond to the two independent solutions of \eqref{ODI2}. However, we have a problem, because 
for $z\in\mathbb{R}_+$ the asymptotic series $\widetilde{\psi}^+$ is not Borel resummable.  Moreover, along the rays  $\phi=\pm 2\pi/3$, where $z = |z| \exp(i \phi)$,  
the argument of the hypergeometric function turns real and the asymptotic series $\widetilde{\psi}^-$ is not Borel resummable.
The solutions $s(\widetilde y_\pm)$ are then not well-defined over the whole complex $z$-plane, but only in wedges delimited by Stokes lines, see fig.\ref{fig:Airy}.
The arrows on the Stokes lines in fig.\ref{fig:Airy} indicate which $\hat \psi_\pm$ is not Borel resummable: $\hat \psi_+$ when the arrow points towards the origin, $\hat \psi_-$ when the arrow points towards infinity.  The formal power series \eqref{ODI4} have a branch-cut singularity at the origin, reported as a curvy red line.
Monodromy considerations are useful to understand that $s(\widetilde y_\pm)$ cannot be globally defined since they are not single-valued around $z=0$. 
Indeed, solutions of the differential equation \eqref{ODI2} should be
single-valued in $z$ around any finite value, including $z=0$, as the only (irregular) singular point of the Airy equation is $z=\infty$. 
We get different functions $y_\pm^{\rm a} = s(\widetilde y_\pm)$, $a={\rm I,II,III}$,  depending on the wedge where $z$ sits,
related with each other through monodromies arising from the Stokes phenomenon. Such monodromies, together with the one arising from the branch-cut singularity, combine to give 
a single-valued solution, as it should be.
A key property of resurgence is that the non-trivial discontinuity of $\widetilde y_+$ is proportional to $\widetilde y_-$ and viceversa.
This phenomenon arises from the fact that the singular behaviour of the Borel functions $\hat \psi_+$ is proportional to $\hat \psi_-$ (and viceversa).
We say that $\hat\psi_-$ ``resurges" from $\hat\psi_+$. In the case of the Airy function, this property boils down to the hypergeometric transformation property
\begin{equation}
\begin{aligned}
{ }_2 F_1(a, b , c, z)=& \frac{\Gamma(c) \Gamma(c-a-b)}{\Gamma(c-a) \Gamma(c-b)}{ }_2 F_1(a, b , a+b+1-c , 1-z) \\
&+\frac{\Gamma(c) \Gamma(a+b-c)}{\Gamma(a) \Gamma(b)}(1-z)^{c-a-b}{ }_2 F_1(c-a, c-b , 1+c-a-b , 1-z)\,,
\label{eq:hypidentity}
\end{aligned}
\end{equation}
\begin{figure}[t!]
    \centering
    \raisebox{-3.em}
 {\scalebox{1.4}{
    \begin{tikzpicture}

   \filldraw[fill=black, draw=black]  (1,3) circle (0.05 cm);
      \draw [stealth-] (2,3) - - (3,3);
            \draw (1,3) - - (2,3);
          \node[right] at (3,3) {\scalebox{.8}{$=S_- $\;\;, }}; 
              \draw [-stealth,red] (2.3,2.5) arc (-30:30:10mm);
        
              \filldraw[fill=black, draw=black]  (5,3) circle (0.05 cm);
                 \draw [-stealth] (5,3) - - (6,3);
            \draw (6,3) - - (7,3);
          \node[right] at (7,3) {\scalebox{.8}{$=S_+ $}}; 
          \draw [-stealth,red] (6.3,2.5) arc (-30:30:10mm);
 
                 \filldraw[fill=black, draw=black]  (9,3) circle (0.05 cm);
              \draw[snake=snake,segment length=5pt] (9,3)-- (11,3); 
          \node[right] at (11,3) {\scalebox{.8}{$=B$}}; 
          \draw [-stealth,red] (10.3,2.5) arc (-30:30:10mm);          
 \end{tikzpicture}}}
    \caption{Connection matrices as we pass an oriented Stokes line (straight lines with arrows) or the branch-cut (wavy lines).}
	\label{fig:Airy2}
\end{figure}
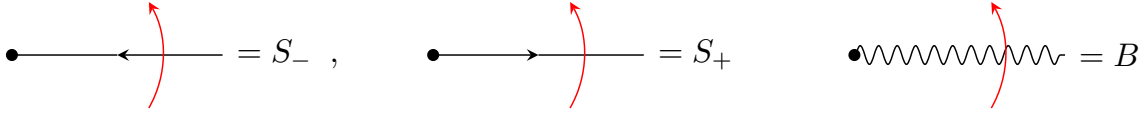 
which shows us that the singular behaviour of ${ }_2 F_1$ at $z=1$ is proportional to the behaviour of another ${ }_2 F_1$ at $z=0$.
Taking $z$ real, the Stokes discontinuity is determined by computing $(s_{\theta^+}-s_{\theta^-})(\widetilde y_\pm)$ for $\theta=0,\pm 2\pi/3$. One has
\begin{align}
y_+^\text{II} - y_+^\text{I} &= i y_-^\text{I} \,, \qquad \qquad y_-^\text{II} - y_-^\text{I}=0  \,,\nn \\
y_+^\text{III} - y_+^\text{II} &=0 \,, \qquad \qquad\;\; y_-^\text{III} - y_-^\text{II} = i y_+^\text{II} \,, \label{ODI7ab} \\
y_+^\text{I} - y_+^\text{III} &=0 \,, \qquad \qquad\;\; y_-^\text{I} - y_-^\text{III} = i y_+^\text{III}  \,. \nn 
\end{align}
It is more convenient not to talk about the jump of the functions $y_\pm$ but of the coefficients of functions linear combinations of $y_\pm$. We write 
\be
y^{\text a}  = c_+^{\text a} y_+ + c_-^{\text a} y_- \,,\qquad a ={\rm I,II,III}\,,
\ee
where $y_\pm$ are the solutions chosen in a given reference wedge, say II.
The connection matrices as we pass a Stokes line or the branch-cut singularity are expressed as $2\times 2$ matrices acting on the coefficients $(c_+^a , c_-^a)$,
see figure \ref{fig:Airy2}:
\be
S_+  =  \left(\begin{array}{cc}
1 & i  \\
0 & 1
\end{array}\right) \,, \qquad 
S_-= \left(\begin{array}{cc}
1 & 0  \\
i & 1
\end{array}\right) \,, \qquad 
B =\left(\begin{array}{cc}
0 & -i  \\
-i & 0
\end{array}\right) \,.
\label{eq:monodromy}
\ee
As anticipated, the total monodromy of $y^a$ under $z\rightarrow e^{2i\pi} z$ is trivial. For example, starting from region I, we have
\be
S_-S_+ B S_+ = I.
\ee
\begin{exr} (recommended)
Prove the first relation in the first row of \eqref{ODI7ab} using the identity \eqref{eq:hypidentity}. In the case at hand, the parameters entering the hypergeometric function satisfy the relation
$a+b=c$ and the second term in \eqref{eq:hypidentity} is obtained in a limit which gives rise to a $\log(1-z)$ term. 
\end{exr}

We are now ready to discuss the solutions of the Airy differential equation \eqref{ODI1} for $x\in \mathbb{R}$. 
There are two independent solutions, commonly denoted ${\rm Ai}(x)$ and ${\rm Bi}(x)$. They respectively decay and blow-up exponentially for $x\rightarrow \infty$.
For ${\rm Ai}(x)$ we have
\be
{\rm Ai}(x) =\left\{\begin{array}{l} y_- \,, \qquad \qquad \;\; x>0 \,, \\
y_-+ i y_+ \,, \qquad x<0 \,.
\end{array}\right. 
\label{eq:AiryA}
\ee
We define the function ${\rm Bi}(x)$ as the orthogonal combination of ${\rm Ai}$ for $x<0$. Its expression at $x>0$ is then given 
by applying $S_+^{-1}$ to its coefficients. We get
\be
{\rm Bi}(x) = \left\{\begin{array}{l} y_+ +  i y_-  \,, \qquad \qquad\; x<0 \,, \\
 2y_+ + i y_- \,, \qquad \qquad x>0 \,.
\end{array}\right. 
\label{eq:AiryB}
\ee

\begin{figure}
\centering
\includegraphics[scale=.4]{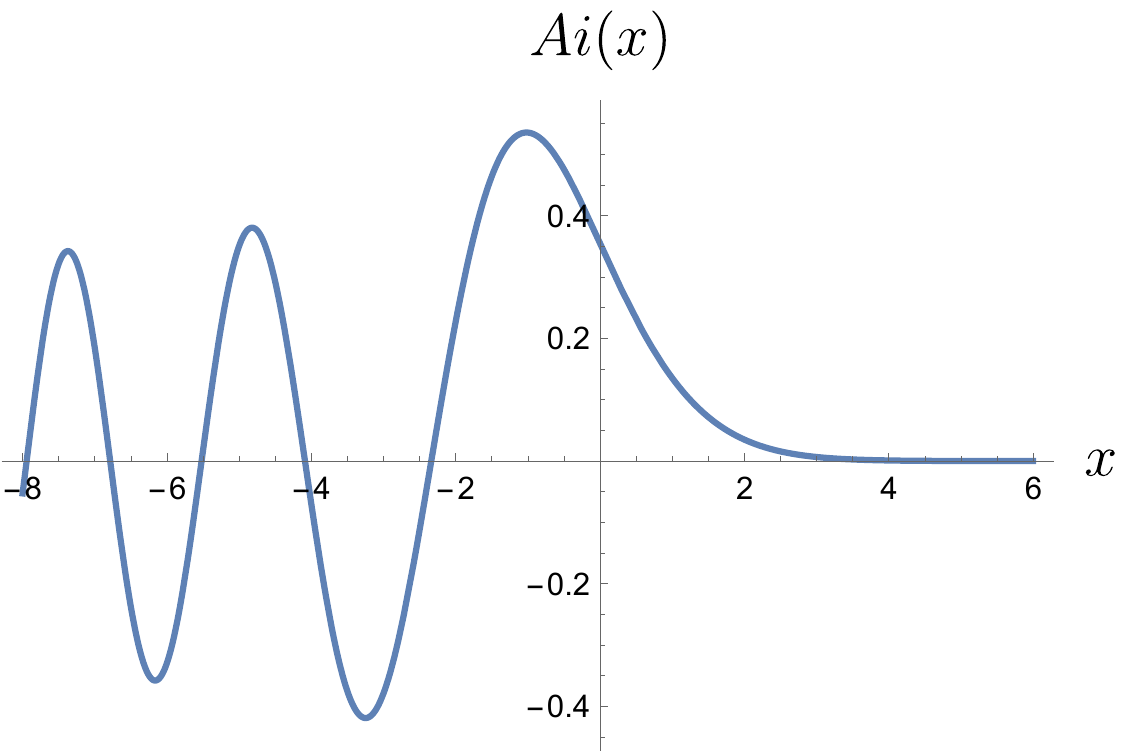}
\caption{\label{fig:AiryAi} The Airy function ${\rm Ai}(x)$ as a function of $x$, oscillatory with a slow power-like decay for $x<0$, and exponentially decaying for $x>0$.}
\end{figure}

Let us summarize the key points we learned:
\begin{enumerate}
\item The same asymptotic series can give rise
to different functions in different wedges of the
complex plane because of the Stokes phenomenon.
\item The ``observables", the functions Ai and Bi, can not always be expressed as a formal asymptotic series, but they can be expressed in terms 
of {\it two} of them.
\item Thanks to the Stokes discontinuity, we can ``discover" one of the two asymptotic series from the non-Borel summability of the other.
\end{enumerate}

We conclude with an historical remark. The Airy function in optics governs the phenomenon of supernumerary rainbows which occasionally can be seen 
accompanying the main bow. The modulus of the peaks of ${\rm Ai}(x)$ describes the intensity
of the different bows, while $x$ is roughly the angular distance from the main bow. Notably, the supernumerary rainbows occur in one direction only with respect to the main bow,
in agreement with the behaviour of ${\rm Ai}(x)$, see figure \ref{fig:AiryAi}. Resurgence in the sky!

\subsection{The Airy function as an integral}

\label{subsec:AiryInt}

For $x>0$, the Airy function ${\rm Ai}(x)$ admits an integral representation of the form
\be
{\rm Ai}(x) = \frac{\lambda^{-\frac 13}}{2\pi} \int_{-\infty}^\infty \! \!dw \, e^{-f(\lambda,w)}\,, \qquad  f=\frac{-i}{\lambda}\Big(\frac{w^3}{3}+w\Big)\,, \qquad x = \lambda^{-\frac 23}\,.
\label{eq:AiryInt1}
\ee
The function $f$ has two extrema at $\pm i$, none of which is along the contour of integration.

In order to apply steepest descent methods for $|\lambda|\ll 1$, we should first understand which saddle point(s) contribute to the integral.
We briefly recap the procedure when $f$ is an arbitrary entire function with isolated and non-degenerate saddle points.\footnote{See e.g. section 3 of \cite{Witten:2010cx} for a more in-depth review, aimed at physicists,  which also features the Airy function as primary example.} 
We first analytically continue $f$ for complex $w$ and deform the initial integration contour ${\cal C}_0$ -- the real $w$ axis in the case of \eqref{eq:AiryInt1} --  into a properly chosen
contour ${\cal C}$ that crosses the saddle points of $f$. In so doing, since we can bring ${\cal C}_0$ up to infinity in some direction in the $w$-plane, ${\cal C}_0$ can effectively split into a sum of  several disjoint contours, i.e. contours connected only at infinity. Let $w_\sigma$ be the saddle points of $f(w)$.
The contour of steepest-descent (also denoted downward flow) ${\cal J}_\sigma$ passing through $w_\sigma$ is determined as the one where ${\rm Im}\,f$ is constant, and Re~$f$ is monotonically {\it increasing} as we leave the fixed point. For each downward flow, there is a dual upward flow ${\cal K}_\sigma$, where ${\rm Im}\,f$ is constant and Re~$f$ is monotonically {\it decreasing} as we leave the fixed point. 

Flows ${\cal J}_\sigma$ which reach Re~$f=\infty$ are called Lefschetz thimbles. Regular upward flows  reach instead Re~$f=\infty$. 
By construction the integral over each Lefschetz thimble is well defined and convergent.
The contour ${\cal C}_0$ should be freely deformed to match a combination ${\cal C}$ 
of  ${\cal J}_\sigma$'s keeping the integral (\ref{eq:AiryInt1}) finite during the deformation. In other words
\be
{\cal C}_0\rightarrow {\cal C} = \sum_\sigma {\cal J}_\sigma n_\sigma\,, \qquad 
\label{eq:AiryInt2}
\ee
where $n_\sigma$ are integers given by the intersection between the {\it original} cycle  ${\cal C}_0$ and the {\it upward} flows ${\cal K}_\sigma$:
\be
n_\sigma = \langle {\cal C}_0, {\cal K}_\sigma \rangle  \,,
\label{eq:AiryInt3}
\ee
where $ \langle a , b \rangle$ denote the intersection pairings between two cycles $a$ and $b$ (with a given orientation). 
Note that, in absence of Stokes lines, downward and upward flows associated to different critical points are dual
to each other:
\be
\langle {\cal J}_\sigma, {\cal K}_\tau \rangle = \delta_{\sigma \tau}\,.
\label{eq:AiryInt4}
\ee

\begin{figure}
\centering
\includegraphics[scale=.4]{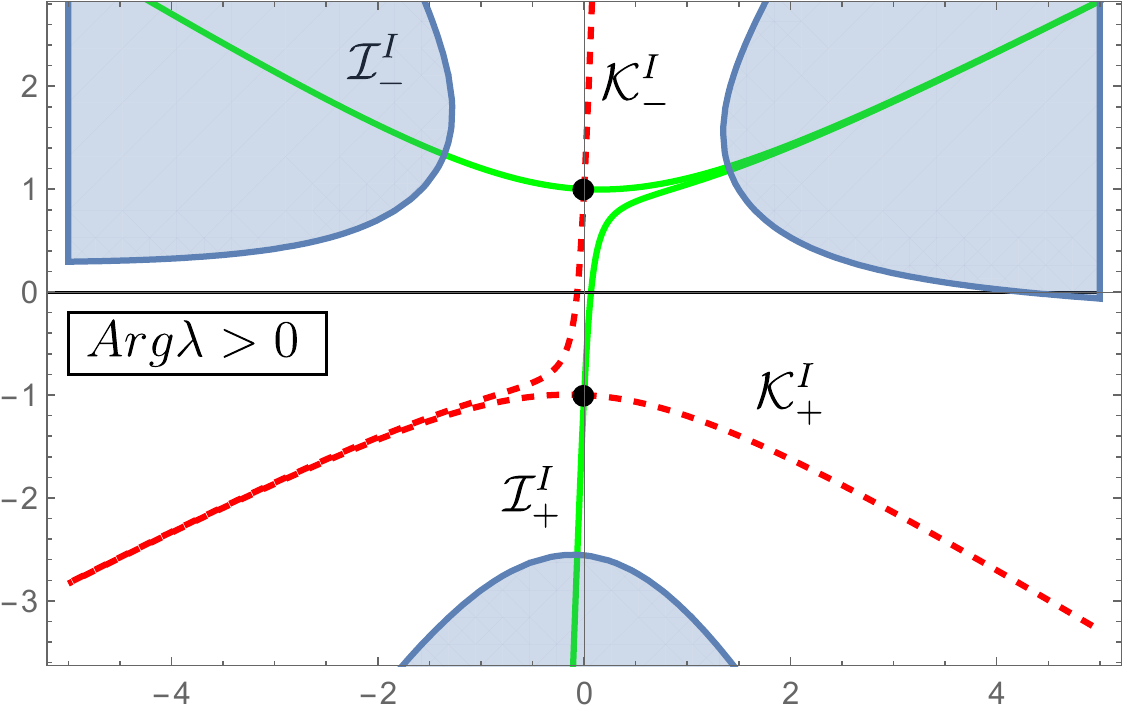}\;\;\;\;
\includegraphics[scale=.4]{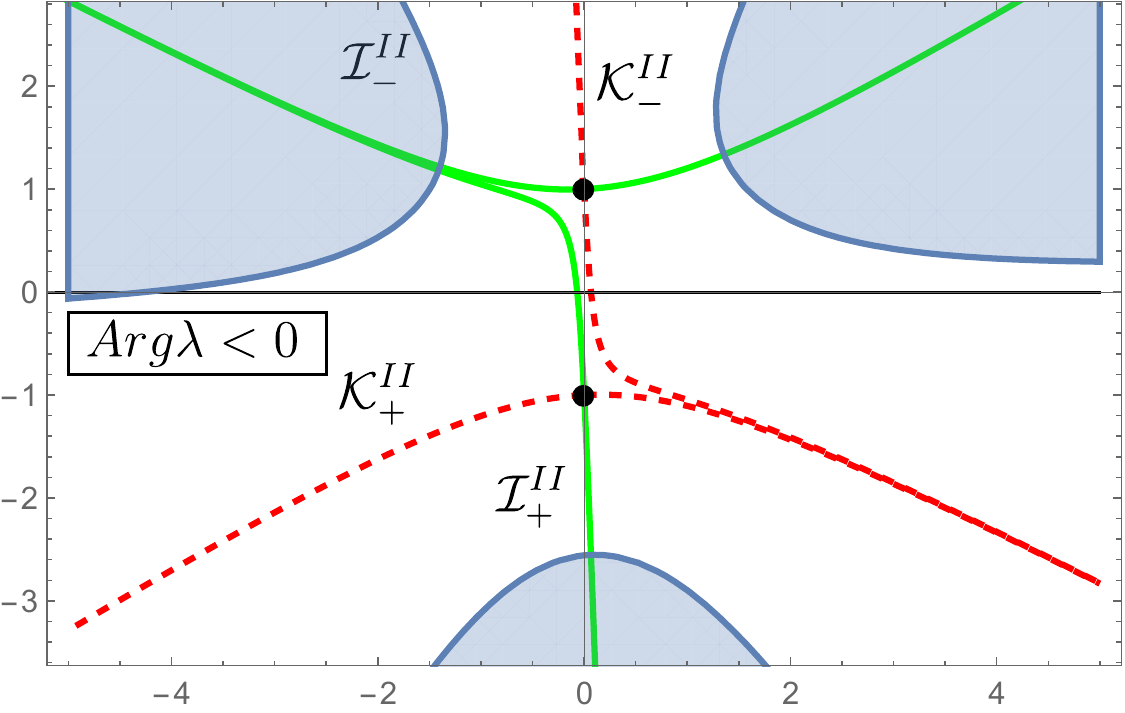}
\caption{Upward (red) and downward (green) flows for the Airy function ${\rm Ai}(x)$ in the $w$-plane. The two black bullets represent the two saddle points at $\pm i$. The shaded blue regions denote the sectors in the complex plane where the integral \eqref{eq:AiryInt1} converges.}
\label{fig:AiryIK} 
\end{figure}

When instead ${\cal J}_\sigma$ hits another saddle point $w_\tau$,\footnote{This can occur if ${\rm Im}\,f(w_\sigma)={\rm Im}\,f(w_\tau)$.}
the flow ends at $w_\tau$ where it reconnects with ${\cal J}_\tau$. This leads to an effective split into two branches and to an ambiguity. 
The flow that starts at $w_\sigma$ and end at $w_\tau$ is denoted a Stokes line. 
In presence of a Stokes line, we  have ${\cal J}_\sigma= {\cal K}_\tau$ and
the corresponding intersection $\langle {\cal J}_\sigma, {\cal K}_\tau \rangle$ 
is not well defined. The problem can be avoided by deforming the initial integral by giving a small imaginary part to $\lambda$, such that all the downward flows turn into Lefschetz thimbles.

We can now come back to the Airy integral \eqref{eq:AiryInt1}. We have $f(w_\pm) = \pm 2/(3\lambda)$, where $w_\pm = \mp i$.\footnote{The reason for this counterintuitive labeling will soon 
be clear.} For $\lambda\in \mathbb{R}$, we have Stokes lines, because
${\cal J}_+= {\cal K}_-$ as ${\rm Im}\, f(w_+)={\rm Im}\, f(w_-)$. The problem is circumvented by assingning a small phase to $\lambda$. We report in figure \ref{fig:AiryIK}
the corresponding downward and upward flows ${\cal J}_\pm$ and ${\cal K}_\pm$. Let us denote by ${\cal J}_\pm^{\rm{I}}$ the downward flows for
${\rm Arg}\,\lambda>0$ and by ${\cal J}_\pm^{\rm{II}}$ the ones for ${\rm Arg}\,\lambda<0$. From figure \ref{fig:AiryIK} we have
\be
\begin{split}
{\cal J}_-^{\rm{II}} & = {\cal J}_-^{\rm{I}} \,, \\
{\cal J}_+^{\rm{II}}  & = {\cal J}_+^{\rm{I}} + {\cal J}_-^{\rm{I}} \,.
\end{split}
\label{eq:AiryInt5}
\ee
The asymptotic expansions around the saddle point $w_\pm$ are in 1-1 correspondence with the formal power series \eqref{ODI4}. We have
\be\begin{split}
& \frac{\lambda^{-\frac 13}}{2\pi} \int_{{\cal I}_-} \! \!dw \, e^{-f(\lambda,w)}\sim  \frac{1}{2\sqrt{\pi}}\lambda^{\frac 16} e^{-\frac{2}{3\lambda} }  \sum_{n=0}^\infty  c_n^{-} \lambda^n = \widetilde y_-(\lambda^{-\frac{2}{3}}) \,, \\
& \frac{\lambda^{-\frac 13}}{2\pi} \int_{{\cal I}_+} \! \!dw \, e^{-f(\lambda,w)}\sim - \frac{i}{2\sqrt{\pi}}\lambda^{\frac 16} e^{\frac{2}{3\lambda} }  \sum_{n=0}^\infty  c_n^{+} \lambda^n =- i \widetilde y_+(\lambda^{-\frac{2}{3}}) \,.
\end{split}
\label{eq:AiryInt6}
\ee
The Stokes discontinuity of $y_+$ along the real line, as given by the first relation in \eqref{ODI7ab} is reproduced by
the discontinuity of ${\cal J}_+$ in \eqref{eq:AiryInt5}, taking into account \eqref{eq:AiryInt6}. 
Independently of ${\rm Arg}\,\lambda$, we have $n_+=0$, $n_-=1$, as ${\cal C}_0$ intersects 
only ${\cal K}_-$. Hence ${\rm Ai}(x)  = s(\widetilde y_-)$, and we reproduce \eqref{eq:AiryA} for $x>0$.

We have established a correspondence between the Stokes discontinuity obtained from considerations of Borel summability and the one coming from integration contours 
in integral representations. The latter perspective makes clear why the singular behaviour of the Borel function $\hat y_+$ is proportional to $\hat y_-$ (and viceversa).
The Airy function has an integral representation also for $x<0$, in which case both saddle points contribute to the integral.

\subsection{Trans-series and bridge equations}

\label{subsec:alien}

Let us come back to the general analysis. It is useful to perform expansions around infinity and redefine the function such that $f(z)\sim z^{-1}$ as $z\rightarrow \infty$.
In this way the formal asymptotic series of $f$, its Borel and generalized Laplace dual transform read
\be
\widetilde f(z) = \sum_{n=0}^\infty f_n z^{-n-1} \,, \quad \hat f(s) = \sum_{n=0}^\infty \frac{f_n s^n}{n!}\,, \quad s_\theta(\widetilde f) = \int_0^{e^{i\theta} \infty}\!\!\!ds \, e^{-s z} \hat f(s)\,, \;\;\;
{\rm Re}\,(e^{i \theta} z)>0\,.
\label{eq:alien1}
\ee
With these conventions, the Borel transforms of $z \widetilde f(z) $ and $\partial_z \widetilde f(z)$ are given as
\be
z \widetilde f(z) \rightarrow \partial_s \hat f(s)\,, \qquad \qquad \partial_z \widetilde f(z) \rightarrow - s \hat f(s)\,.
\label{eq:alien2}
\ee
A Gevrey-1 formal series is said to be {\it resurgent} if its Borel transform has endless analytic continuation, namely no natural boundary of singularities.
A resurgent function is called {\it simple} if
\be
\hat f(s) = \frac{c'_\omega}{2i\pi(s-\omega)} +\frac{c_\omega}{2i\pi} \hat \psi(s-\omega)\log (s-\omega) + {\rm reg.}\,,
\label{eq:alien3}
\ee
close to a singular point at $s=\omega$. In \eqref{eq:alien3}, reg. denotes terms which are analytic at $s=\omega$ and $\hat\psi$ is analytic at $s=\omega$. In general, 
the nature of the singularity of a Borel function is not of logarithmic type. If this is the case, one may use \eqref{eq:basic10} to possibly find the values of $b$
for which \eqref{eq:alien3} applies. This step will not be needed in what follows. In the Airy function example described in section \ref{subsec:AiryDE}, the Borel functions were simple, with  $c'_\omega=0$.
 
Given a Borel function $\hat f(s)$, we denote by $\omega_a= |\omega_a| \exp(i \theta)$ the set of all its singularities along a given ray in the Borel plane. The 
singularities induces a Stokes discontinuity \eqref{eq:resurgence4} as we pass the ray with angle $\theta$. In order to disentangle the individual contribution of a singularity,
we (indirectly) define the alien derivative $\Delta_{\omega_a}$ as 
\be
\mathfrak{S}_{\theta}\equiv \exp\Big(\sum_{\{\omega_a\}} e^{-z \omega_a} \Delta_{\omega_a} \Big)\,.
\label{eq:alien4}
\ee
It is also useful to define the dotted alien derivative
\be
\dot \Delta_\omega \equiv e^{-\omega z} \Delta_\omega
\label{eq:alien5}
\ee
and the total alien dotted derivative 
\be
\dot \Delta_\theta = \sum_{\{\omega_a\}} \dot \Delta_{\omega_a} = \log \mathfrak{S}_{\theta}\,.
\label{eq:alien6}
\ee
Let us work out the action of the alien derivative on a formal power series in the case in which we have only one singularity $\omega$ along the ray $\theta = {\rm Arg}\,\omega$.
For simplicity we set $c_\omega'=0$ in \eqref{eq:alien3}. We compute
\be
(s_+-s_-)_\theta \widetilde f =\int_\omega^{e^{i \theta\infty}} \!\! \!\! ds\, e^{-s z} c_\omega \hat \psi(s-\omega) = \int_0^{e^{i \theta\infty}} \!\! \!\!du\, e^{-(\omega+u) z} c_\omega \hat \psi(u)
= c_\omega e^{-\omega z} s_{\theta}(\widetilde \psi)\,,
\label{eq:alien7}
\ee
where $\widetilde \psi$ is the asymptotic series associated to the Borel function $\hat \psi$ appearing in \eqref{eq:alien3}, which we assumed for simplicity to be 
Borel resummable along the ray $\theta$. Using \eqref{eq:resurgence3} and \eqref{eq:resurgence4} we have
\be
\mathfrak{S}_{\theta}(\widetilde f) = \widetilde f +  c_\omega e^{-\omega z} \widetilde \psi\,.
\label{eq:alien8}
\ee
Matching with \eqref{eq:alien4} gives
\be
\Delta_\omega  \widetilde f = c_\omega \widetilde \psi\,, \qquad \Delta_\omega^n  \widetilde f =0\,, \;\; n>1\,.
\label{eq:alien9}
\ee
Whenever $\widetilde \psi$ is not Borel resummable, higher alien derivatives do not vanish. We will see an example of this kind in what follows.
The alien derivative is a map between formal power series. Given a series $\widetilde f$, the alien derivative extracts the power series associated to the 
singular behaviour of $\hat f$, in this case $\widetilde \psi$.  
It is called a derivative because it satisfies the Leibniz rule
\be
\Delta_\omega (\widetilde f_1 \widetilde f_2) = \Delta_\omega(\widetilde f_1) \widetilde f_2 + \widetilde f_1 \Delta_\omega(\widetilde f_2) \,.
\label{eq:alien10}
\ee
The alien derivative does not commute with the ordinary derivative $\partial_z$. However, using the correspondence \eqref{eq:alien2}, it is 
easy to show that its dotted version does:
\be
[\dot \Delta_\omega,\partial_z ] = 0\,.
\label{eq:alien11}
\ee
The action of the alien derivatives can also be extracted using so called bridge equations.
We describe them in a particularly simple set-up in terms of differential equations \cite{Dorigoni:2014hea,abs}, the minimum needed for the subsequent application in the context of integrable models we will consider. See \cite{sauzin2014} for a more extensive and math-oriented discussion.

Let $\Phi(z)$ be the solution of a differential equation in $z$. We look for a perturbative formal power series solution in $1/z$ in terms of a trans-series of the form
\be\label{eq:alien12}
\widetilde \Phi(z,\sigma) = \sum_{n=0}^\infty \sigma^n e^{-\omega_n z} \widetilde \phi_n(z)\,.
\ee
In \eqref{eq:alien12} $\widetilde \phi_0$ is the perturbative formal power series, $\widetilde \phi_{n>0}$ are the non-perturbative ones, and 
we have taken the Stokes parameters $\sigma_n = \sigma^n$, with $\sigma$ to be determined. By definition the trans-series $\widetilde \Phi(z,\sigma)$ satisfies the differential equation to all orders in $1/z$
and $e^{-z}$.  Suppose also that the differential equation is such that $\partial_\sigma \widetilde \Phi$ and $\dot \Delta_{\omega_n} \widetilde \Phi$ satisfy the
same (homogeneous) differential equation.\footnote{For example, given that $[\partial_\sigma,\partial_z]=0$ and \eqref{eq:alien11}, this condition is automatically satisfied
for non-linear inhomogeneous differential equations which are first order in $\partial_z$, the inhomogeneous term being killed by either the action of $\partial_\sigma$
or $\dot\Delta_{\omega_n}$.} Then we have the equation
\be\label{eq:alien13}
\dot \Delta_{\omega_k} \widetilde \Phi(z,\sigma) =  A_k(\sigma)  \partial_\sigma \widetilde \Phi(z,\sigma) \,,
\ee
where $A_k$ are undetermined functions of $\sigma$. The relation \eqref{eq:alien13} is an example of bridge equation, so called because it allows to relate
different formal power series $\widetilde \phi_n$ between each other. Let us assume that $\omega_n = 2n$, with $n> 0$, and let us write $\dot \Delta_{\omega_n} \equiv \dot \Delta_n$ for simplicity.
We have
\be\label{eq:alien14}
\begin{split}
\dot \Delta_k \widetilde \Phi(z,\sigma) & = \sum_{n=0}^\infty \sigma^n e^{-2(n+k) z} \Delta_k \widetilde \phi_n(z) \,, \\
A_k(\sigma) \partial_\sigma  \widetilde \Phi(z,\sigma) & = \sum_{n=0}^\infty n \sigma^{n-1} A_k(\sigma) e^{-2n z} \widetilde \phi_n(z) \,.
\end{split}\ee
Matching powers of $\sigma$ and $e^{-2n z}$ gives
\be
A_1(\sigma) = A_1\,, \qquad \qquad  A_{k>1}(\sigma) =0\,,
\ee
and
\be\label{eq:alien15}
\Delta_1 \widetilde \phi_n = (n+1) A_1 \widetilde \phi_{n+1} \,,
\ee
with $A_1$ a constant coefficient. This can be determined by looking at the behaviour of the Borel function $\hat \phi_0$ associated to the perturbative series $\widetilde \phi_0$ 
around the singularity $\omega_1$.  Once $A_1$ is known, the resurgent structure of the trans-series is essentially determined. However, we still have to understand how to get the actual exact function $\Phi(z)$, solution of the starting differential equation (with some boundary conditions), from the trans-series $\widetilde \Phi(z,\sigma)$. Given \eqref{eq:alien15}, the Stokes automorphism $\mathfrak{S}$ reads
\be\label{eq:alien16}
\mathfrak{S}_0  \widetilde \Phi(z,\sigma) = e^{\dot \Delta_1} \widetilde \Phi(z,\sigma) = e^{A_1 \partial_\sigma} \widetilde \Phi(z,\sigma) =  \widetilde \Phi(z,\sigma+A_1) \,,
\ee
namely
\be\label{eq:alien17}
s_+\big( \widetilde \Phi(z,\sigma)\big) = s_-(\big(  \widetilde \Phi(z,\sigma+A_1)\big)\,,
\ee
where $s_\pm$ here means Borel resummations of the individual formal power series $\widetilde \phi_n$. We still have to fix $\sigma$.
When the exact function $\Phi(z)$ is real, we can determine $\sigma$ and reconstructs $\Phi$ from $\widetilde \Phi$ demanding the reality 
of the resummation procedure. Indeed, both $s_+$ and $s_-$ will in general gives rise to complex values for $\Phi$. 
It is useful to introduce the median Borel resummation defined as
\be\label{eq:alien19}
s_{{\rm med}} = s_+ \mathfrak{S}^{-\frac 12} = s_-  \mathfrak{S}^{\frac 12} \,.
\ee
In contrast to $s_\pm$, $s_{{\rm med}}$ preserves the reality properties of the formal power series where it acts on.
If $C$ denotes complex conjugation, one has $C s_+ = s_- C$ and
\be
C \mathfrak{S}_0 = \mathfrak{S}_0^{-1} C\,, \qquad\qquad \qquad C \dot \Delta_0 = - \dot \Delta_0 C.
\ee 
Therefore, from \eqref{eq:alien19}, we get
\be
C s_{{\rm med}} = s_{{\rm med}} C\,.
\ee
Remarkably, the resummation of the whole trans-series is reproduced starting from the real perturbative series by using median resummation. In our case we have\footnote{We tacitly assume here that the sum of formal power series defining the trans-series is absolutely convergent. As far as we are aware, a proof of this statement is missing. See e.g. \cite{costin2001,Marino:2023epd} for results pointing towards this direction.}  
\begin{align}
s_{{\rm med}}(\widetilde \phi_0)  & = s_+ \Big(e^{-\frac 12 e^{-2z} \Delta_1} \widetilde \phi_0\Big) = 
s_+ \bigg(\sum_{l=0}^\infty \Big(-\frac 12\Big)^l\frac{e^{-2zl}}{l!} \Delta_1^l \widetilde \phi_0 \bigg) = 
\sum_{l=0}^\infty \Big(-\frac{A_1}{2}\Big)^l e^{-2zl} s_+(\widetilde \phi_l) \nn \\
& = s_- \Big(e^{\frac 12 e^{-2z} \Delta_1} \widetilde \phi_0\Big) = 
s_- \bigg(\sum_{l=0}^\infty \Big(\frac 12\Big)^l\frac{e^{-2zl}}{l!} \Delta_1^l \widetilde \phi_0 \bigg) = 
\sum_{l=0}^\infty \Big(\frac{A_1}{2}\Big)^l e^{-2zl} s_-(\widetilde \phi_l)\,,
\label{eq:alien20}
\end{align}
where we have used 
\be
\Delta_1^l \widetilde \phi_0 = A_1^l l! \widetilde \phi_l\,, 
\ee
see \eqref{eq:alien15}.
The trans-series in the first and second row of \eqref{eq:alien20} matches with \eqref{eq:alien12} and \eqref{eq:alien17}, with the identification
\be
\sigma = \sigma_+\equiv  -\frac{A_1}{2}, \qquad \sigma + A_1 = \sigma_-\equiv  \frac{A_1}{2}\,.
\ee
We reconstruct the exact function by considering 
\be\label{eq:alien18}
s_{{\rm med}}(\widetilde \phi_0)  = s_+\big(\widetilde \Phi(z,\sigma_+)\big) = s_-\big( (\widetilde \Phi(z,\sigma_-)  \big) = \Phi(z)\,.
\ee
Given the trans-series $\widetilde \Phi(z,\sigma)$, the choice of Stokes constant $\sigma$ depends on which lateral resummation we perform in such a way
that a well-defined non-ambiguous result is obtained. This example concretely shows how from the knowledge of the perturbative series $\widetilde \phi_0$ it is possible to reconstruct the full result using 
resurgence techniques. The trans-series we will determine in the next section for the large $N$ free energy in the principal chiral field model is
precisely of this form.

\section{$2d$ models and $1/N$ expansion}

\label{sec:2d1N}

The $1/N$ expansion is a time-honored tool to detect phenomena which are otherwise invisible in a coupling constant expansion, such as the qualitative emergence of spontaneous chiral 
symmetry breaking in four-dimensional  gauge theories \cite{coleman-witten} or its quantitative description  in certain two-dimensional models such as the Gross-Neveu theory \cite{gross-neveu}. 
This is possible because $1/N$ rearranges the perturbative series and resums it order by order in $1/N$.
In an ordinary perturbation theory with coupling constant $g$, at fixed $N$ we have an expansion of the form
\be\label{eq:2dN1}
\widetilde \phi_0(g,N)  = \sum_{n=0}^\infty c_n(N) g^n\,, \qquad c_n(N)  \underset{n\gg 1}{\sim} n! \,.
\ee
At large $N$, we define the 't Hooft coupling $\lambda = g^2 N$, kept fixed for $g\rightarrow 0$ and $N\rightarrow \infty$, and write the expansion
\be\label{eq:2dN2}
\widetilde \psi_0(g,N)  = \sum_{n=0}^\infty d_n(\lambda) N^{-n}\,.
\ee
In the large $N$ case \eqref{eq:2dN2}, two large order behaviours can be considered:
\begin{enumerate}
\item fixed $\lambda$, behaviour of $1/N$ series. 
\item fixed $n$, behaviour of expansion of $d_n(\lambda)$ for $\lambda\ll 1$:  $\widetilde d_n(\lambda)=\sum_{k=0}^\infty d_{n,k} \lambda^k$. 
\end{enumerate}
In case 1 we generally expect 
\be\label{ral, ineq:2dN3}
d_n(\lambda)  \underset{n\gg 1}{\sim} n!\,, \qquad \text{fixed} \, \lambda\,,
\ee
though there are situations where the large $N$ series can converge for specific observables, e.g. the free energy in the 2d Gross-Neveu model \cite{DiPietro:2021yxb}. 
The qualitative behaviour of $\widetilde d_n$ in case 2 depends on the theory. 
Since instantons are suppressed at large $N$, the factorial growth of Feynman diagrams is reduced to exponential at each order in $1/N$.
As a consequence, in theories without renormalons, $\widetilde d_n$ is a convergent series in $\lambda$. Notable theories of this kind include zero-dimensional (0d) matrix models, 4d ${\cal N}=4$ super Yang--Mills theory, $3d$ $O(N)$ models or Chern--Simons--matter theories. When renormalons are present, however, the series $\widetilde d_n$ is divergent asymptotic, with
\be\label{eq:2dN4}
d_{n,k}  \underset{k\gg 1}{\sim} k! \,,\qquad \text{fixed} \, n\,.
\ee
We consider in what follows case 2, where we analyze the first orders in $1/N$ and look at the behaviour of the series for $d_n(\lambda)$ in theories where this is divergent due to the presence of renormalons.  An interesting class of theories of this sort are 2d integrable QFTs like the $O(N)$ non-linear sigma model (NLSM), the $O(N)$ Gross-Neveu (GN) model, the $SU(N)$ principal chiral field (PCF) model. All such theories are gapped in the IR, they have a dynamically generated scale, they are UV free. The first two are vector theories, the third is a matrix model. 
We mostly focus on the last case.

In this third and final lecture we compute the $1/N$ leading order free energy in presence of a chemical potential, 
and show how the result could have been obtained
from a perturbative series in the t'Hooft coupling constant, using median resummation.

We review in section \ref{subsec:Fh} how the free energy can be computed using TBA. The explicit computation for the principal chiral field at leading order in $1/N$ is carried out in sections \ref{subsec:FhPCF} and \ref{subsec:TSF0}.  Results for other models, at large and finite $N$, are briefly discussed in section \ref{subsec:other}.

\subsection{Free energy in integrable systems}
\label{subsec:Fh}

The observable of interest is the relative free energy
\be
{\cal F}(h) \equiv  F(h)-F(0)\,,
\label{eq:FhPCF1}
\ee
where $F(h)$ is the zero temperature and infinite volume limit of the free energy in presence of a chemical potential $h$ associated to a conserved charge $Q$.
Per unit volume, $F(h)$ reads
\be
\label{eq:FhPCF2}
F(h) =-\lim_{V, \beta \rightarrow \infty} {1\over V\beta } \log \, {\rm tr} \, e^{-\beta (H-h Q)}, 
\ee
where $V$ is the volume of space, $H$ is the Hamiltonian, and $\beta$ is the total length of Euclidean time. 
For simplicity, we refer to ${\cal F}(h)$ just as the free energy. 
  
It was pointed out in \cite{pw} that in integrable quantum field theories ${\cal F}(h)$ can be determined by using the TBA ansatz and exact $S$-matrix 
amplitudes, in terms of a linear integral equation. Let $m$ be the mass gap in the theory. For $h>m$,  we expect that the ground state will be populated with particles
with positive charge with respect to $Q$. By appropriately embedding the charge operator $Q$ within the full global symmetry of the theory, 
the ground state will be populated by a single species of particle, the one minimizing $H-h Q$, in the limit of zero temperature.
The number density $\rho$ of such particles can be determined in terms of Bethe roots $\chi(\theta)$ by the TBA equation 
\be
\label{eq:FhPCF3}
\chi(\theta)-\int_{-B}^B  d \theta' \, K(\theta-\theta') \chi (\theta')=m \cosh \theta\,, 
\ee
where $\chi(\theta)$ is supported on the interval $[-B,B]$, with $B$ a quantity yet to be determined.
The integral kernel appearing in the Bethe ansatz equation is given by 
\be
\label{eq:FhPCF4}
K(\theta)=\frac{1}{2 i\pi} \frac{d  \log S(\theta)}{d\theta},
\ee
where $\theta=\theta_1-\theta_2$ is the relative rapidity of the two scattering states, and $S$ is their corresponding $S$-matrix amplitude, exactly determined
using integrability \cite{zz}. Independently of their spin, the particles which populate the ground state obey a Fermi statistics.\footnote{In the bosonic NLSM and PCF theories this
is seen from the fact that $S(0) = -1$. In the GN model $S(0)=1$, but the particles in the vacuum are fermions.} 
The energy per unit length $e$ and the density $\rho$ are given by
\be\label{eq:FhPCF5}
e={m \over 2 \pi} \int_{-B}^B d \theta \, \chi(\theta)  \cosh \theta, \qquad \rho={1\over 2 \pi} \int_{-B}^B d \theta\, \chi(\theta). 
\ee
The parameter $B$ is related to the Fermi momentum $p_F$ of particles. Its value is fixed by the density $\rho$ and one can obtain an equation of state relating $e$ to $\rho$. 

The free energy ${\cal F}(h)$ is obtained by taking a Legendre transform of $e(\rho)$:
\be\label{eq:FhPCF6}
\begin{split}
h  &\equiv \partial_\rho e(\rho) \,, \quad \quad \;\;\; {\cal F}(h)  \equiv e(\rho) - \rho h\,, \\
\rho & = - \partial_h {\cal F}(h) \,, \quad \quad  e(\rho)  = {\cal F}(h)  + \rho h\,.
\end{split}
\ee
In an equivalent, more useful,  formulation of the TBA equations, the basic quantity is a function $\epsilon (\theta)$, which satisfies the integral equation 
\be
\label{eq:FhPCF7}
\epsilon(\theta) -\int_{-B}^{B} d\theta'\, K(\theta-\theta') \epsilon(\theta')=h-m \cosh \theta\,,
\ee
and the boundary conditions
\be
\epsilon(\pm B)=0\,.
\label{eq:FhPCF8}
\ee
In terms of $\epsilon$, the free energy is more directly given by
\be
\label{eq:FhPCF9}
{\cal F}(h)=  -{m \over 2 \pi}\int_{-B}^B \!\! d \theta  \, \cosh \theta \,  \epsilon(\theta)\,.
\ee
In order to gain some intuition on the physical interpretation of the above formulas, it is useful to work out the case in which the scattering is negligible and we have a gas of free fermions.
In this case, \eqref{eq:FhPCF3} gives $\chi=m \cosh\theta = dp/d\theta$, where $p = m \sinh \theta$ is the momentum of the particles. The density is given by
\be\label{eq:FhPCF9a}
 \rho={m\over \pi} \sinh B =  {1\over 2 \pi} \int_{-p_F}^{p_F} d p = \frac{p_F}{\pi}\,,
 \ee
while the energy
\be\label{eq:FhPCF9b}
 e={1\over 2 \pi} \int_{-p_F}^{p_F} \sqrt{p^2+m^2} d p  \,.
 \ee
The function $\epsilon$ reads
\be\label{eq:FhPCF9c}
\epsilon(\theta) = h - m \cosh \theta \,.
\ee
Given \eqref{eq:FhPCF8} and \eqref{eq:FhPCF9a}, we have 
\be\label{eq:FhPCF9d}
\cosh B = \frac{h}{m} = \sqrt{1+\frac{p_F^2}{m^2}}\,, 
\ee
and hence
\be\label{eq:FhPCF9e}
\epsilon(p) = \sqrt{p_F^2+m^2} -  \sqrt{p^2+m^2}\,.
\ee
For $p<p_F$ ($|\theta|< B)$ we can interpret $\epsilon$ as the energy of hole excitations, configurations where a particle of momentum $p$ is removed from the Dirac sea.
On the other hand, for $p>p_F$ ($|\theta|> B)$, $\epsilon$ can be seen (up to a sign) as the energy of a particle excitation. 
 \begin{exr}
Show that 
\be 
{\cal F}(h) =  \underset{\rho}{\rm min}\; e(\rho)  - \rho h 
\ee
corresponds to ${\cal F}(h)$ defined in \eqref{eq:FhPCF9},  where $\epsilon(\theta)$ satisfies the condition \eqref{eq:FhPCF8}.
\end{exr}
Computed in this way, ${\cal F}(h)$ is a well-defined function of $h$ and $m$. No renormalons appear. In order to recast the result in terms of asymptotic expansions 
in a QFT perturbative series, we have to ``reinsert"  the dependence on the coupling constant $g^2$ which appears in the UV Lagrangian description of such theories.
A useful definition of running coupling is
\be
\frac{1}{\alpha(\mu)} + \xi  \log \alpha(\mu) \equiv \log\Big( \frac{\mu}{m} \Big)\,,
\label{eq:FhPCF10}
\ee
where 
\be
\alpha \equiv 2 \beta_0 g^2\,,\qquad \xi = \frac{\beta_1}{2 \beta_0^2} \,, 
\label{eq:FhPCF11}
\ee
and $\beta_0$, $\beta_1$ are the one- and two-loop coefficients of the beta-function for $g$ defined as
\be
\label{eq:FhPCF11a}
\beta(g)= \mu \frac{d g}{d \mu} =-\beta_0 g^3 - \beta_1 g^5 + {\cal O}(g^7)\,.
\ee
In terms of $\alpha$, we have 
\be
\label{eq:FhPCF12}
\beta(\alpha)= \mu {d \alpha \over d \mu} =-\alpha^2 - \xi \alpha^3 + {\cal O}(\alpha^4)\,.
\ee
We call TBA scheme the renormalization scheme defined by \eqref{eq:FhPCF10}.
Applying $\mu \partial_\mu$ to \eqref{eq:FhPCF10} gives the exact $\beta$-function in the TBA scheme.
Consistently, the first two terms in the expansion agree with those in \eqref{eq:FhPCF12}.  

The considerations above apply to the NLSM, GN and PCF model. In what follows we focus on the PCF model, see
\cite{DiPietro:2021yxb} for details concerning the NLSM and GN theories. 
We can expand ${\cal F}(h)$ in $1/N$:
\be\label{eq:FhPCF13}
{\cal F}(h) \approx \sum_{k=0}^\infty {\cal F}_k(h) \Delta^{k-1}\,, \quad \quad \Delta = \frac 1N \,.
\ee
Determining the explicit forms of ${\cal F}_k$ is not an easy task, so we consider only the leading term ${\cal F}_0$. 
In the PCF model it is convenient to define the 't Hooft  coupling
\be\label{eq:FhPCF14}
\alpha \equiv \alpha \Big(\mu=\sqrt{\frac{2\pi}{e}} h\Big)\,,
\ee
where $\alpha(\mu)$ is the TBA coupling defined in \eqref{eq:FhPCF10}, and
expand ${\cal F}_0(h)$ in terms of it. We have
\be\label{eq:FhPCF15}
{\cal F}_0(h) \sim -\frac{h^2}{8\pi} \widetilde{\Phi}_{k=0}(\alpha,\sigma_\ell)\,,
\ee
where
\be
\widetilde \Phi_k(\alpha,\sigma_\ell) =   \sum_{\ell=0}^\infty \sigma_{\ell} e^{-\frac{2\ell}{\alpha}}  \widetilde\phi_k^{(\ell)}(\alpha)\,,
\label{eq:FhPCF16}
\ee
where $\sigma_{\ell}$ are the trans-series parameters introduced in \eqref{eq:resurgence4b}.
We compute the exact form of ${\cal F}_0$ in section \ref{subsec:FhPCF} and determine the associated trans-series $\widetilde \Phi_0$ in section \ref{subsec:TSF0}.

\subsection{${\cal F}(h)$ in the principal chiral field at leading order in $1/N$}
\label{subsec:FhPCF}

The PCF model is a matrix quantum field theory. Its Lagrangian density reads
\be\label{eq:FEC1}
{\cal L}= {1\over  g^2} {\rm tr} \left(\partial_\mu \Sigma\,  \partial^\mu \Sigma^\dagger\right)\,,
\ee
where $\Sigma$ is a $SU(N)$-valued matrix field. The coupling $g^2$ is UV free and the theory is strongly coupled in the IR.
The $\beta$-function parameters defined in \eqref{eq:FhPCF11} read
\be
\beta_0 = \frac{N}{16\pi}\,, \qquad \qquad \xi =\frac  12\,.
\ee
The full mass spectrum of the theory has been determined using integrability \cite{Wiegmann:1984ec}.
We have a set of bound states with masses
\be\label{eq:FEC2}
m_n = m \frac{\sin\Big(\frac{\pi n}{N}\Big)}{\sin\Big(\frac{\pi}{N}\Big)}\,, \qquad n=1, 2,\ldots, N-1\,,
\ee
in the rank $n$ antisymmetric representation $(n,n)$ of the $SU(N)_L\times SU(N)_R$ global symmetry group of \eqref{eq:FEC1}.

In the PCF model we embed $Q$ in $SU(N)_V \subset SU(N)_L \times SU(N)_R$.
Its eigenvalues in the fundamental representation are denoted by $\boldsymbol{q}=(q_1, \cdots, q_N)$.   We choose the embedding \cite{pcf}\footnote{See \cite{fkw1,fkw2,Kazakov:2023imu} for other charge assignments.}
\be
\label{or-charges}
 \boldsymbol{q}= \left({1\over 2}, -{1 \over 2(N-1)}, \cdots, -{1\over 2(N-1)} \right). 
 \ee
 There is only one particle state minimizing $H-h Q$, which is given by the $(11)$ entry of the states in the bifundamental representation of $SU(N)_L\times SU(N)_R$.
 In the scattering of such states only the symmetric channel (in both $SU(N)_L$ and $SU(N)_R$) of the $S$-matrix can contribute. This is given by 
\cite{Wiegmann:1984ec}
\be\label{eq:FhPCF2a}
S(\theta) = -\frac{\Gamma\Big(1+\frac{i \theta}{2\pi}\Big)^2\Gamma\Big(\frac 1N-\frac{i \theta}{2\pi}\Big)\Gamma\Big(1-\frac 1N-\frac{i \theta}{2\pi}\Big)}{\Gamma\Big(1-\frac{i \theta}{2\pi}\Big)^2
\Gamma\Big(\frac 1N+\frac{i \theta}{2\pi}\Big)\Gamma\Big(1-\frac 1N+\frac{i \theta}{2\pi}\Big)}\,.
\ee 

We compute ${\cal F}_0 (h)$ by expanding in $1/N$ the TBA equation \eqref{eq:FhPCF7}. 
This expansion of the kernel is non-analytic in $1/N$ and gives: 
\be\label{eq:FhPCF17}
K(\theta) = \delta(\theta) + \Delta K_1(\theta) + {\cal O}(\Delta^2)\,, 
\ee
where
\be\label{eq:FhPCF18}
K_1(\theta) = - \frac{d}{d\theta}\Big(\frac{2}{\theta}\Big)\,.
\ee
We also expand $\epsilon$ and $B$:
\be\label{eq:FhPCF19}
\epsilon(\theta) = \Delta^{-1} \sum_{k=0}^\infty \epsilon_k(\theta)\Delta^k \,, \qquad B=\sum_{k=0}^\infty B_k \Delta^k\,.
\ee
The ${\cal O}(\Delta^{-1})$ trivializes, while at ${\cal O}(\Delta^0)$ we get
\be\label{eq:FhPCF20}
- \int_{-B_0}^{B_0} \! d\theta'\,  K_1(\theta-\theta')\epsilon_0(\theta') = h - m \cosh \theta\,.
\ee
Relabeling $B_0\rightarrow B$, using the explicit form \eqref{eq:FhPCF18} of $K_1$ and integrating over $\theta$, \eqref{eq:FhPCF20} is equivalent to 
\be\label{eq:FhPCF20a}
{\rm P}  \int_{-B}^{B} \! d\theta' \frac{2}{\theta-\theta'}\epsilon_0(\theta') = V'(\theta) \,, \qquad V'(\theta) \equiv h \theta - m \sinh \theta\,.
\ee
Interestingly enough, \eqref{eq:FhPCF20a} is identical to the equation for the density of eigenvalues of a one-matrix model \cite{Brezin:1977sv} with potential 
\be\label{eq:FhPCF21}
V = \frac{h}{2} \theta^2 - m \cosh \theta\,,
\ee
where $\theta$ and $\epsilon_0(\theta)$ play the role of the eigenvalues and their density, respectively. We can then adapt the matrix model solutions in terms of resolvents
to our case. Let us briefly recall how resolvents work. We look for an analytic function $R(z)$ in the $z$-cut complex plane, with a cut $[-B,B]$ over the real axis, of the form
\be\label{eq:FhPCF22}
R(z) = \int_{-B}^B \! dx\, \frac{2\epsilon_0(x)}{z-x}\,.
\ee
By construction, the discontinuity along the cut of the resolvent gives $\epsilon_0$, and we require that its principal part equals $V'$:
\be\label{eq:FhPCF23}
\begin{split}
R(z^+) - R(z^-) & = - 4i \pi \epsilon_0(z) \,, \\
\frac 12 \big(R(z^+) + R(z^-) \big)&  = V'(z)\,.
\end{split}
\ee
Thanks to its analyticity properties, $R(z)$ can be written as 
\be\label{eq:FhPCF24}
R(z) = V'(z) - \sqrt{z^2-B^2} F_1(z)\,,
\ee
where 
\be\label{eq:FhPCF25}
\epsilon_0(z) = \frac{1}{2\pi}\sqrt{B^2-z^2} F_1(z)\,,
\ee
 and 
 \be\label{eq:FhPCF26}
 F_1(z) = \oint_{C_\infty} \frac{dw}{2i \pi} \frac{V'(w)}{w-z} (w^2-B^2)^{-\frac 12}= \oint_{C_0} \frac{dw}{2i \pi} \frac{V'(\frac 1w)}{1-w z} (1-w^2 B^2)^{-\frac 12}
 \,.
 \ee
 Expanding both $ V'(w^{-1})$ and $(1-w^2 B^2)^{-1/2}$ around $w=0$ and using the relation
  \be\label{eq:FhPCF27}
\oint_{C_0} \frac{dw}{2i \pi} (1-z w)^{-1} w^{-s} =  \left\{\begin{array}{l} z^{s-1}  \,, \quad\; s\geq 0 \,, \\
0 \,, \qquad \quad s<0 \,,
\end{array}\right. 
 \ee
 gives
  \be\label{eq:FhPCF28}
 \epsilon_0(\theta)= {1\over 2 \pi} {\sqrt{B^2-\theta^2}} \left( h- 2m \sum_{r,k \ge 0} {2k \choose k} {r+ k+1 \over (2(r+k+1))!} \left( {B^2 \over 4}\right)^k \theta^{2r} \right). 
 \ee
Thanks to the overall $\sqrt{B^2-\theta^2}$ factor, the boundary condition \eqref{eq:FhPCF8} is automatically satisfied by \eqref{eq:FhPCF28}.
However, due to the form of $K_1$ in \eqref{eq:FhPCF18}, we have to also demand regularity of $\partial_\theta \epsilon_0$ at $\theta = \pm B$. This condition gives
 \be\label{eq:FhPCF29}
 h- 2 m \sum_{r,k \ge 0} {2k \choose k} {r+ k+1 \over (2(r+k+1))!} 4^{-k} B^{2(k+r)}=0\,.
 \ee
 The series can be resummed, leading to 
\be\label{eq:FhPCF29}
 \frac{h}m = \frac 2B I_1(B) + I_2(B)  \,,
 \ee
where $I_k$ are Bessel functions of the first kind. The relation \eqref{eq:FhPCF29} implicitly defines $B$ as a function of $m/h$.
Knowing $\epsilon_0$, we can determine the free energy \eqref{eq:FhPCF9} by using 
\be\label{eq:FhPCF30}
\int_{-B}^B\! d\theta \,\sqrt{B^2-\theta^2} \theta^{2k} = B^{2k+2}\frac{\sqrt{\pi}\Gamma\big(k+\frac 12\big)}{2\Gamma(k+2)}\,.
\ee
Eventually we get
\be
\label{eq:FhPCF31}
{\cal F}_0(h)= -{h^2\over 4 \pi} \left\{ {B^2 I_1(B) \over 2 I_1(B ) +B  I_2(B) }- 
{1\over 2} {B^4 \over  \left( 2 I_1(B) +B  I_2(B) \right)^2} \, {}_1 F_2\left( {1\over 2} ; 1, 2; B^2 \right)  \right\}\,,
\ee
where $B$ is determined from \eqref{eq:FhPCF29} and ${}_1 F_2$ is the generalized hypergeometric function (not to be confused with the ordinary hypergeometric function
${}_2 F_1$)
\be
\label{eq:FhPCF32}
 {}_1 F_2(a; b_1, b_2; z) = \sum_{k=0}^\infty \frac{(a)_k z^k}{(b_1)_k (b_2)_k k!} \,.
\ee

\subsection{Trans-series expansion of ${\cal F}_0$}
\label{subsec:TSF0}

The free energy  \eqref{eq:FhPCF31} is an exact expression, function of the two mass scales in the problem, $h$ and $m$. 
The next task is to rewrite \eqref{eq:FhPCF31} as a trans-series in terms of the coupling defined in \eqref{eq:FhPCF10}.
The task is greatly simplified by the fact that both the Bessel functions $I_a$ and the  generalized hypergeometric  ${}_1 F_2$
admit simple trans-series expansions in terms of their argument. 

The Bessel functions of the second kind $I_a(z)$ and $K_a(z)$ are solutions of the second order differential equation $z^2 y''+z y'-(z^2+a^2)y=0$.
Formal solutions of the equation in terms of asymptotic power series in $1/z$ can be found, similarly to the case of the Airy equation discussed in section \ref{subsec:AiryDE}.
These are given by
\be\label{eq:TS1}
\widetilde y_\pm^{(a)} =\sqrt{\frac{\pi}{2z}} e^{\pm z} \sum_{n=0}^\infty c_n^\pm (a) z^{-n}\,,
\ee
where
\be\label{eq:TS2}
c_n^\pm(a) = (\pm 1)^n \frac{\Gamma\Big(n+\frac 12 + a\Big)\Gamma\Big(n+\frac 12 - a\Big)}{2^n \Gamma(n+1) \Gamma\Big(\frac 12 + a\Big) \Gamma\Big(\frac 12 - a\Big)}\,.
\ee
For real $z$, $\widetilde y_-^{(a)}$ is Borel resummable and its Borel resummation defines the Bessel function $K_a$, while $\widetilde y_+^{(a)}$ has a Stokes discontinuity:
$(s_+ - s_-)(\widetilde y_+^{(a)}) =  2i \cos(\pi a) s(\widetilde y_-^{(a)})$. The Bessel function $I_a$ is obtained  as the resummation of a two-term trans-series:
\be\label{eq:TS3}
\pi I_a(z) = s_\pm(\widetilde y_+^{(a)}) + \sigma_\pm e^{\mp i \pi a} s(\widetilde y_-^{(a)})\,, \qquad \sigma_\pm = \mp i \,.
\ee
By using \eqref{eq:TS1} and \eqref{eq:TS3}, we get the following equalities, valid for $B>0$:
\begin{align}
{2 \over B} I_1(B)+ I_2(B) & = {e^B \over {\sqrt{2 \pi B}}} \left( s_\pm (\widetilde \gamma^{(0)})+ \sigma_\pm e^{-2B} s(\widetilde \gamma^{(1)}) \right)\,, \nn \\
 B I_1(B) & =  {\sqrt{ B \over 2 \pi}} e^B \left(s_\pm(\widetilde \nu^{(0)}) - \sigma_\pm e^{-2B} s(\widetilde \nu^{(1)}) \right), 
\label{eq:TS4}
\end{align}
where
\begin{align}
\widetilde \gamma^{(0)}(B)& =1+ {1\over 8 B}+ {9 \over 128 B^2}+ \cdots, \qquad\widetilde \gamma^{(1)}(B)= \widetilde\gamma^{(0)}(-B) \,, \nn \\
\widetilde \nu^{(0)}(B)& = 1-{3 \over 8B} -{15 \over 128 B^2}+ \cdots, \qquad \widetilde\nu^{(1)}(B)= \nu^{(0)}(-B)\,,
\label{eq:TS5}
\end{align}
are Gevrey-1 series.  

A similar analysis for the generalized hypergeometric function gives
 \be
 \label{eq:TS6}
 {B^2 \over 2} {}_1 F_2\left( {1\over 2} ; 1, 2; B_0^2 \right) ={ e^{2B} \over 4 \pi} \left( s_\pm ( \widetilde f^{(0)})+ C^{\pm} e^{-2B} 4B s_\pm ( \widetilde f^{(1)})+ 
 e^{-4B}  s( \widetilde f^{(2)}) \right), 
 \ee
 where
 \be
  \label{eq:TS7}
 \widetilde f^{(0)}(B)=1+{1\over  4B}+ \cdots, \qquad \widetilde  f^{(2)}(B)= f^{(0)}(-B), \qquad  \widetilde f^{(1)}(B)= 1-{1\over 8 B^2}+ \cdots
 \ee
 are again Gevrey-1 asymptotic series. As before, the symbol $s$ with no subscript refers to formal series which have a non-ambiguous Borel resummation along the positive real axis.
Using \eqref{eq:TS4} and \eqref{eq:TS6} we can determine a trans-series expansion of the free energy \eqref{eq:FhPCF31} in terms of $1/B$ and $e^{-2B}$.
We are however interested in obtaining the trans-series expansion in terms of the 't Hooft coupling $\alpha$ introduced in (\ref{eq:FhPCF10}), 
which is more directly connected to conventional perturbation theory. The relation between $\alpha$ and $B$ is obtained from \eqref{eq:FhPCF29} and the first relation in \eqref{eq:TS4}.
With $\alpha$ defined as in \eqref{eq:FhPCF14}, we have
\be \label{eq:TS8}
B-{1\over 2} \log(B) -{1\over 2}+ \log  \widetilde \gamma^{(0)}(B)+ 
\log \left(1+ \sigma_\pm  e^{-2B} { \widetilde\gamma^{(1)}(B) \over  \widetilde\gamma^{(0)}(B)} \right) = {1\over  \alpha}+ {1\over 2} \log ( \alpha), 
\ee
and it has a trans-series solution of the form 
\be \label{eq:TS9}
B= {1\over \alpha} \widetilde{\mathfrak{B}}^{(0)}(\alpha)+ \sum_{\ell \ge 1} \sigma_\pm ^\ell e^{-{2 \ell /  \alpha}} \widetilde{\mathfrak{B}}^{(\ell)}(\alpha)\,,
\ee
where the series  $\widetilde{\mathfrak{B}}^{(\ell)}(\alpha)$ can be computed at arbitrarily high order (see \cite{DiPietro:2021yxb} for the explicit form of the first terms for $\ell=0,1,2$). 
Plugging \eqref{eq:TS9} in the trans-series in $1/B$ for ${\cal F}_0$ finally gives us the desired result of a trans-series structure of ${\cal F}_0(h)$ in $\alpha$ and $\exp(-2/\alpha)$:
\be
\label{eq:TS10}
\begin{aligned}
\widetilde \Phi_0(\alpha,\sigma_\pm)=  & {1\over  \alpha}- {1\over 2} - { \alpha\over 4} -{5 \alpha^2 \over 16} -{53 \alpha^3 \over 96}-\frac{487 \alpha ^4}{384}-\frac{13789 \alpha ^5}{3840}-\frac{185143 \alpha ^6}{15360}+{\cal O}\left(\alpha
   ^7\right)\\
&-{4 \sigma_\pm \over e   \alpha^2} e^{-2 /\alpha} \left( 1 + \alpha + {\alpha^2 \over 4} -\frac{\alpha ^3}{16}+\frac{\alpha ^4}{96}-\frac{31 \alpha ^5}{384}-\frac{23 \alpha ^6}{1280}+{\cal O}\left(\alpha ^7\right)  \right)\\
&+{2 \sigma^2_\pm\over e^2  \alpha} e^{-4 / \alpha} \left( 1-\frac{\alpha }{4}+\frac{3 \alpha ^2}{8}-\frac{\alpha ^3}{2}+\frac{4 \alpha ^4}{3}-\frac{181 \alpha ^5}{64}+\frac{3227 \alpha
   ^6}{320}+{\cal O}\left(\alpha ^7\right)\right) \\
   &+ {\cal O} \left( e^{-6 / \alpha}  \right)\,,
\end{aligned}
\ee
where $\widetilde \Phi_0$ is defined in \eqref{eq:FhPCF15}. We see that the trans-series \eqref{eq:TS10} is of the form given in \eqref{eq:FhPCF16}, where
the parameters $\sigma_\ell$ are all related and given by
\be
\sigma_\ell = \sigma_\pm^\ell\,,
\ee
with a two-fold ambiguity $\pm$ deriving from the choice of lateral Borel resummation.
The first terms of the first series $\widetilde \phi_\ell$ read
\begin{align}
\widetilde \phi_0(\alpha)& = \frac{1}{\alpha}-{1 \over 2} -\frac \alpha 4 +\cdots, \nn  \\
\widetilde\phi_1(\alpha) & = -{4 \over e \alpha^2 }\left( 1 + \alpha + \frac{\alpha^2}{4} + \cdots\right) \,, \label{eq:TS11} \\
\widetilde\phi_2(\alpha) & = {2 \over e^2 \alpha}\left( 1 -{ \alpha \over 4} + \frac{3\alpha^2}{8} + \cdots\right). \nn
\end{align} 
The non-perturbative corrections in \eqref{eq:TS10} are proportional to powers of $\exp(-2/\alpha)$, which corresponds to $\Lambda^2$, the square of the dynamically generated scale.
From the considerations in section \ref{subsec:reno} it is natural to identify the Borel singularities associated to the non-perturbative $\exp(-2\ell /\alpha)$ terms to IR renormalons
and to the appearance of condensates with UV scaling dimensions $2\ell$.\footnote{In vector models this interpretation is confirmed by large $N$ techniques which allow us to compute the values of such condensates.} Despite we get an infinite number of IR renormalons, the trans-series $\widetilde \Phi_0$ inherits the simple resurgent properties of its building blocks, the Bessel functions $I_{1,2}$ and the generalized hypergeometric function ${}_1 F_2$. In fact, an analysis of the first formal series $\widetilde\phi_n$ points toward a very simple pattern of alien derivatives:\footnote{Similar resurgent equations  apply to the solution to Painlev\'e II equations \cite{Marino:2008ya}.}
\be \label{eq:TS12}
\Delta_1 \widetilde\phi_n = 2i (n +1) \widetilde\phi_{n+1} \,.
\ee
This is precisely of the same form as \eqref{eq:alien15}, with $A_1 = 2i$, and $A_{k>1}=0$.\footnote{It would be nice to derive \eqref{eq:TS12} directly from some bridge equations, as 
in the toy example discussed in section \ref{subsec:alien}.}
This allows us to use the results discussed at the end of section \ref{subsec:alien}.
More specifically, we have
\be\label{eq:TS13}\boxed{
{\cal F}_0(h) = -\frac{h^2}{8\pi} s_\pm(\widetilde{\Phi}_{0}(\alpha,\sigma_\pm))= - \frac{h^2}{8\pi} s_{\rm med}(\widetilde{\phi}_{0}(\alpha))\,,}
\ee
where $s_{\rm med}$ is the median resummation defined in \eqref{eq:alien19}. See \cite{DiPietro:2021yxb} for numerical checks of the validity of \eqref{eq:TS13}.

Note that $\widetilde{\phi}_{0}$ corresponds to the asymptotic  series around the perturbative wrong vacuum  with massless excitations. 
The IR renormalons can be seen as a signal of the inconsistency of the expansion. Yet, and quite remarkably,  \eqref{eq:TS13} shows that by the knowledge
of $\widetilde \phi_0$, we could have in principle recovered the full exact result. This, of course, would have required to work out in a bottom-up approach
 the pattern of alien derivatives \eqref{eq:TS12}, and assume that median resummation gives the exact result.

\subsection{Other large $N$ and finite $N$ results}
\label{subsec:other}

We briefly discuss in this last section other results obtained both at large and finite $N$. 
Most of the analysis have focused  on the  $O(N)$ NLSM, the $O(N)$ GN model, and the $SU(N)$ PCF model.\footnote{See \cite{Schepers:2023dqk} for an exception.}
By using a technique developed in \cite{volin-thesis,volin} to extract many perturbative terms in $1/B$ or in the coupling $\alpha$ from 
the TBA equations \eqref{eq:FhPCF3} and \eqref{eq:FhPCF7}, \cite{mr-ren} established that the large order behaviour in the above theories are controlled by renormalons,
for any finite $N$. Evidence for identifying such singularities with renormalons was provided from large $N$ QFT computations around the ``naive" vacuum, which was shown to give rise
to non-Borel resummable asymptotic series with the expected singularity structure \cite{mmr}.
While in the PCF model resurgence works beautifully at large $N$, as synthesised in \eqref{eq:TS13}, 
for the $O(N)$ non-linear and Gross-Neveu vector models it was found that, when fixing the order in the $1/N$ expansion, the resulting perturbative series in the coupling constant is not sufficiently generic and cannot be used to predict non-perturbative corrections \cite{DiPietro:2021yxb}.  Interesting results were found for the $O(N)$ NLSM at $N=4$, where technical simplifications allow for a more in-depth analysis.  In particular, exploiting the results of \cite{volin-thesis,volin}, \cite{abbh1,abbh2,Bajnok:2021dri} found resurgent relations between the perturbative and the first few non-perturbative series. 
An efficient way to extract the first terms of the non-perturbative series using Wiener-Hopf techniques was found in \cite{Marino:2021dzn}.
Leveraging on this method, \cite{Bajnok:2022xgx} found a way to determine the whole trans-series starting from the perturbative series, and pointed out that 
the exact result is recovered by median resummation of the trans-series.  The $O(N)$ model with $N=3$ also received some interest, being notably the only model with 
topologically stable instantons. The trans-series in this case is given by both renormalon and instantons singularities and its structure has been
analyzed in \cite{Bajnok:2021zjm,Bajnok:2022rtu,Marino:2022ykm}. The resurgence properties of the $O(N)$ NLSMs have been
fully worked out recently in \cite{Bajnok:2024qro}, where it has been found that median resummation of the perturbative series is enough to reconstruct the whole result for $N\geq 4$,
while for $N=3$ at least two other non-perturbative series are required. 
Evidence that, in contrast to the large $N$ case, the perturbative series might be enough to predict non-perturbative corrections in the GN model at finite $N$ has been given in \cite{Marino:2021dzn}. In addition, new  Borel singularities were found which are naively not those expected from ordinary renormalons, i.e. integer powers of the dynamically generated scale. 
The physical interpretation of these singularities requires further study.

\section{Concluding remarks}
\label{conclusions}

Perturbative expansions in QFT typically have zero radius of convergence,
yet this does not harm its application at weak coupling, its
natural area of application. 
Somewhat surprisingly, perturbation theory can also be used at strong coupling.
Before resurgence, strong coupling was accessible only for Borel
resummable expansions, a very stringent constraint. Several results have been obtained
in this way, mostly in the context of critical phenomena in scalar field theories. 

Non-Borel resummable perturbative expansions can be systematically studied by using resurgence.
We still don't know the general requirements under which perturbative expansions in QFT are resurgent, so
it is useful to provide evidence in theories where we know the exact answer by other means. 
We have shown in these lectures the notable case of the free energy as a function of a chemical potential
in 2d integrable models. 

So far we mostly checked that resurgence works in
theories where we knew the answer by other means.
So the big question is: can we use it to actually compute new observables?
In QFT this seems to be at the moment out of reach, because we need to know 
many more perturbative terms than those typically known in order to ``activate" the resurgence program.
In quantum mechanics it has been shown that, using suitable deformations, observables
expressed in terms of a trans-series in ordinary perturbation theory also admit expansions which
give rise to a single Borel resummable asymptotic series \cite{power}.  
And history has shown that predictions can
be made when we have to deal with a single series only. 
Understanding how resurgence works in controlled cases would be of help in possibly
finding similar deformations in QFT. 

Even in absence of a method to get quantitative results, resurgence
can be seen as a useful organizing principle for strongly
coupled theories, since the size of non-perturbative effects is efficiently and analytically captured. 

\section*{Acknowledgments}

I would like to thank Lorenzo Di Pietro, Marcos Mari\~no and Giacomo Sberveglieri for collaboration on \cite{DiPietro:2021yxb}, and
Tomas Reis for discussions on related topics and comments on the manuscript.
These notes are an expanded version of lectures given at IHES, and review talks given at a workshop in Les Diablerets and a colloquium at Mainz, in 2023. 
I would like to thank IHES, and especially Slava Rychkov, for hospitality, the organizers of the Les Diablerets workshop Marcos Mari\~no  and Ricardo Schiappa,
and Tobias Hurth for the invitation to Mainz. Work partially supported by INFN Iniziativa Specifica ST\&FI.

\bibliographystyle{JHEP}
\bibliography{Refs}

\end{document}